%% file: manuscript_SIM.tex
\documentclass[a4paper,12pt]{article}

\usepackage{natbib}
\usepackage{graphicx}
\usepackage{amsmath}
\usepackage{authblk}
\usepackage{color}
\usepackage{float}
\usepackage{caption}
\usepackage{subcaption}
\usepackage[hidelinks]{hyperref}
\usepackage{multirow} 
\usepackage[labelformat=simple]{subcaption} 
\usepackage[margin=1.1in]{geometry}

\hypersetup{
  colorlinks = true, 
  urlcolor = blue, 
  linkcolor = black, 
  citecolor = black 
}

\begin{document}
	
	\title{A maximum penalised likelihood approach for semiparametric accelerated failure time models with time-varying covariates and partly interval censoring}

	\author[1]{Aishwarya Bhaskaran}
	\author[1,2]{Ding Ma}
	\author[1,3]{Benoit Liquet}
 	\author[4,5,6,7]{Angela Hong}
	\author[8]{Stephane Heritier}
        \author[4,5,9]{Serigne N Lo}
	\author[1]{Jun Ma}
	\affil[1]{School of Mathematical and Physical Sciences, Macquarie University, Sydney, NSW, Australia}
        \affil[2]{ULTRA Team, Centre for Clinical Research, The University of Queensland, Brisbane, QLD, Australia}
        \affil[3]{Laboratoire de Mathématiques et de leurs Applications, Université de Pau et des Pays de l'Adour, E2S UPPA, CNRS, Anglet, France}
        \affil[4]{Melanoma Institute Australia, The University of Sydney, Sydney, NSW, Australia}
        \affil[5]{Faculty of Health and Medicine, The University of Sydney, Sydney, NSW, Australia}
        \affil[6]{Department of Radiation Oncology, Chris O’Brien Lifehouse, Sydney, NSW, Australia}
        \affil[7]{GenesisCare, Radiation Oncology, Mater Sydney Hospital, Sydney, NSW, Australia}
	\affil[8]{School of Public Health and Preventive Medicine, Monash University, Melbourne, VIC, Australia}
        \affil[9]{Charles Perkins Centre, The University of Sydney, Sydney, NSW, Australia}

    \date{\today}
	
    \maketitle 
    \newpage
 
	\begin{abstract}
		Accelerated failure time (AFT) models are frequently used to model survival data, providing a direct quantification of the relationship between event times and covariates. These models allow for the acceleration or deceleration of failure times through a multiplicative factor that accounts for the effect of covariates. While existing literature provides numerous methods for fitting AFT models with time-fixed covariates, adapting these approaches to scenarios involving both time-varying covariates and partly interval-censored data remains challenging. Motivated by a randomised clinical trial dataset on advanced melanoma patients, we propose a maximum penalised likelihood approach for fitting a semiparametric AFT model to survival data with partly interval-censored failure times. This method also accommodates both time-fixed and time-varying covariates.  We utilise Gaussian basis functions to construct a smooth approximation of the nonparametric baseline hazard and fit the model using a constrained optimisation approach. The effectiveness of our method is demonstrated through extensive simulations. Finally, we illustrate the relevance of our approach by applying it to a dataset from a randomised clinical trial involving patients with advanced melanoma.
	\end{abstract}

	
        \section{Introduction}
        Cox proportional hazards models \citep{Cox1972} and accelerated failure time (AFT) models \citep{Prentice1978} serve as two of the leading approaches to modelling covariate effects on survival data. The use of an AFT model is appealing as it asserts a direct relationship between the time-to-event and covariates, wherein the failure times are either accelerated or decelerated by a multiplicative factor in the presence of these covariates. Furthermore, the AFT model emerges as a viable alternative when the proportional hazards assumption, required to use a Cox model, does not hold.
         
        To avoid specifying the error distribution or the corresponding baseline hazard function, one may choose to employ a semiparametric AFT model. However, this may complicate the estimation of regression coefficients compared to a parametric AFT model, particularly in cases where the model includes interval-censored survival times \citep{Li2020} or time-varying covariates \citep{Ma2023}. Currently, there are several model fitting methods available in the literature for fitting semiparametric AFT models with time-fixed covariates. These include rank-based estimator approaches \citep{Prentice1978, Jin2003, Chiou2014}, the Buckley-James (least-squares type) estimator approach \citep{Buckley1979, Jin2006} and its generalised version \citep{Gao2017}, a smoothed error distribution approximation approach \citep{Komarek2005} and a baseline hazard approximation approach \citep{Li2020}. However, extending most of these methods to settings involving both time-varying covariates and interval censoring poses significant challenges.
     
        Time-varying covariates are essentially covariates whose values change over time. In biomedical research, examples of such covariates include biomarkers and measures of cumulative exposure to treatments over time. The values of these covariates are usually collected over time at pre-determined or random intermittent time points that occur prior to the censoring or failure time. It is also important to consider that, in some cases, the monitoring of event status and the tracking of time-varying covariates may be unrelated, particularly when diagnostic methods for the event of interest are costly or invasive. Ultimately, these covariates may heavily influence the time-to-event and thus the inclusion of time-varying covariates in survival models is pertinent. To handle time-varying covariates in an AFT model, \cite{Zeng2007} proposed an approach that involves constructing a kernel-smoothed approximation to the profile likelihood function to allow for gradient-based optimisation and estimation of the regression parameters. However, this method is limited to right-censored data.
    
    In practice, time-to-event data often involve complex censoring mechanisms, necessitating methods that can handle various types of censoring. For instance, partly interval-censored survival data, as defined by \cite{Kim03}, include exact event times and potentially left, right and interval censoring times as part of the observed survival data. Consider, for example, the treatment of long-term illnesses or medical studies that require periodic follow-up appointments with clinicians. At each of these visits, the event status is monitored, which often leads to the occurrence of interval-censored data. Given the prevalence of interval censoring in practice, it is important to account for this type of data when developing a survival regression model. A commonly used strategy of replacing a censoring interval with a single imputation, such as the middle-point, is not ideal as it may introduce biases in the estimated regression coefficients \citep{Ma2021, Webb2023}. Multiple imputation can be employed to handle interval censoring for the Cox model \citep{Pan00}; however, its computational demands are substantial for AFT models with time-varying covariates because it requires numerous model fittings, each of which may be time-consuming, rendering the process impractical. 

    Our work in this paper is motivated by the WBRTMel trial, a multinational, open-label, phase III randomised controlled trial that compared whole brain radiotherapy (WBRT) to observation (control). In this study, patients were randomised to either the WBRT arm or the control arm following the first local treatment (surgical excision and/or stereotactic irradiation treatment) of 1 to 3 melanoma brain metastases. The primary outcome was the proportion of patients experiencing distant intracranial failure, as determined by magnetic resonance imaging, within 12 months of randomisation. The initial study concluded that adjuvant WBRT did not provide evidence of clinical benefit with respect to the primary outcome \citep{Hong2019}. This result was based on the non-statistically significant effect of WBRT on the primary outcome, assessed using logistic regression without covariate adjustment, as specified in the statistical analysis plan \citep{Lo2019}. 
    
    Patients who were alive without intracranial progression at the trial's closure were right-censored. However, for patients who experienced the primary outcome, intracranial failure was not observed exactly but occurred between two scan visits, resulting in partly interval-censored data. Additionally, some patients received systemic therapy during the study, which is a time-varying covariate influencing progression. Given the efficacy of systemic therapy in extending the time to progression \citep{long2018combination}, it is crucial to account for this time-varying covariate when assessing the effectiveness of WBRT. Thus, in response to the need for a methodology that accommodates both time-varying covariates and interval censoring in AFT models, this paper aims to develop an efficient approach for fitting AFT models with partly interval censoring while incorporating time-varying covariates.
	 
    Building on the foundations laid by \cite{Li2020} and \cite{Ma2023}, we construct a smooth approximation of the unknown baseline hazard by using Gaussian basis functions, rather than the kernel-smoothed profile likelihood approach of \cite{Zeng2007}. The number of basis functions is determined based on the sample size of the dataset. We then alternate between implementing a pseudo-Newton method and the multiplicative iterative (MI) algorithm \citep{ChMa12} to carry out maximum penalised likelihood (MPL) estimation. The pseudo-Newton step updates the regression coefficients, while the MI step updates the Gaussian basis coefficients. Here, imposing a penalty term allows one to enforce an additional layer of smoothness,  complementing the Gaussian basis functions, in the approximation of the baseline hazard function. It also provides greater flexibility in the estimation of the baseline hazard as the solutions become less dependent on the number and placement of basis functions. As demonstrated in the simulation study in Section \ref{sec:simulation}, this approach yields small biases in the regression coefficients for both time-fixed and time-varying covariates across different sample sizes. Moreover, the mean estimated baseline hazard functions closely align with the true baseline functions.
	 
	 The remainder of this paper is organised as follows. In Section \ref{sec:notation}, we introduce the notation required for this paper. Section \ref{sec:AFTmodel} formally presents the model definition for a semiparametric AFT model with time-varying covariates. In Section \ref{sec:MPL}, we discuss how maximum penalised likelihood estimation can be carried out to estimate the regression coefficients and baseline hazard. We also discuss readily available asymptotic results, particularly for the case when active constraints are present. The results from the simulation studies and the application to a real data example are reported in Sections \ref{sec:simulation} and \ref{sec:application} respectively. Finally, we conclude with the relevance of the new method and highlight some potential research directions in Section \ref{sec:discussion}.
	 
    \section{Notation}\label{sec:notation}
    For an individual $i$, where $ 1 \leq i \leq n$, let $T_i$ denote the event time of interest or failure time. Under a partly interval censoring scheme, each individual is associated with a random vector $\boldsymbol{Y_i} = [Y_i^L, Y_i^R]^T$ such that $T_i \in [Y_i^L, Y_i^R]$ and $Y_i^L \leq Y_i^R$. The observed value of this random vector is denoted as $\boldsymbol{y_i} = [y_i^L, y_i^R]^T$.

     Throughout this article, we assume independent censoring given the covariates;
     see, for example, Chapter 1 of \cite{Sun06}. Let $\delta_i$ denote the indicator variable corresponding to the observed failure time and let $\delta_i^L, \delta_i^R$ and $\delta_i^I$ denote the indicator variables corresponding to left, right and interval censoring respectively. These indicator variables are introduced to help simplify the likelihood expression. Subsequently, each $\boldsymbol{y_i}$ either corresponds to an event time ($y_i^L = y_i^R = t_i, \delta_i = 1$), left censoring time ($y_i^L = 0, \delta_i^L = 1$), right censoring time ($y_i^R = \infty, \delta_i^R = 1$) or interval censoring time ($y_i^L < y_i^R, \delta_i^I = 1$). 
	
	Each individual can have both time-fixed and time-varying covariates. For time-fixed covariates, the values are denoted by a $1 \times p$ vector $\boldsymbol{x}_i = (x_{i1},\dots, x_{ip})$. For time-varying covariates, at each time point $t$, the values are represented by a $1\times q$ vector $\boldsymbol{z}_i(t) = (z_{i1}(t), \dots, z_{iq}(t))$. Let $\tilde{\boldsymbol{z}}_i(t)$ denote the history of $\boldsymbol{z}_i(s)$ up to time $t$, namely  $\tilde{\boldsymbol{z}}_i(t) = \{\boldsymbol{z}_i(s): s \leq t\}$. The observations we consider are $(\boldsymbol{y}_i, \delta_i, \delta_i^L, \delta_i^R, \delta_i^I, \boldsymbol{x}_i, \tilde{\boldsymbol{z}}_i(y_i^*))$, for $1\leq i \leq n$, where we assume $y_i^* = y_i^LI(y_i^R=\infty)+y_i^RI(y_i^R\neq \infty)$ with $I(\cdot)$ denoting an indicator function.
	
 	In practice, time-varying covariates $z_{ir}(t)$, where $1 \leq r \leq q$, are rarely given as continuous functions of $t$. Instead, under an intermittent sampling scheme, the values of $\boldsymbol{z}_i(t)$ are only observed at specific time points $t_{i1}, \dots, t_{in_i}$, where we define $t_{i1} = 0$, allowing for the possibility that different individuals have distinct sampling schedules for their time-varying covariates. We assume, without loss of generality, that when $T_i$ is right censored, $t_{in_i} = y_i^L$; otherwise, $t_{in_i} = y_i^R$. However, note that the event status assessment times may not necessarily align with the sampling schedules for the time-varying covariates. 

    Now let $\boldsymbol{z}_i(t_{ia}) = (z_{i1}(t_{ia}), \dots, z_{iq}(t_{ia}))$ correspond to the vector of time-varying covariates observed at a particular sampling point $t_{ia}$ for an individual $i$. Then, the observed values for the time-varying covariates across all sampling points for an individual $i$ can be represented by the following $n_i \times q$ matrix: $\boldsymbol{Z}_i = (\boldsymbol{z}_i(t_{i1})^T, \dots, \boldsymbol{z}_i(t_{in_i})^T)^T$. Combining these $\boldsymbol{Z}_i$ matrices together, a $N \times q$ matrix is formed as follows
	\begin{equation*}
		\boldsymbol{Z} = (\boldsymbol{Z}_1^T, \dots, \boldsymbol{Z}_n^T)^T,
	\end{equation*}
	where $N = \sum^n_{i=1} n_i$.
  
    Table \ref{table:tab1} allows us to visualise the long data format of such time-varying covariates, a format widely used when dealing with time-varying covariates in Cox models. The entries in the ``Status" column indicate whether or not the event of interest may have occurred within the corresponding time interval. In the example, individual 1 is left-censored while individual 2 is right-censored. 
	
	\begin{table}[ht]
        \caption{An example of a snippet of time-varying data with different sampling points.}
		\label{table:tab1}
		\centering
		\begin{tabular}{lllllll}
			\hline
			Individual & Start  & End & Status & $z_{i1}(t)$ & \dots & $z_{iq}(t)$\\
			\hline
            1 & $t_{11}$ & $y_1^R$ & 1 & $z_{11}(t_{11})$ & \dots & $z_{1q}(t_{11})$ \\\\
            2 & $t_{21}$ & $t_{22}$ & 0 & $z_{21}(t_{21})$ & \dots & $z_{2q}(t_{21})$ \\
            \phantom{1} & $t_{22}$ & $y_2^L$ & 0 & $z_{21}(t_{22})$ & \dots & $z_{2q}(t_{22})$ \\\\
            \vdots &  $\cdots$ & $\cdots$ & $\cdots$ & $\cdots$ & $\cdots$ & $\cdots$\\
			\hline
		\end{tabular}
	\end{table}

	\section{Semiparametric AFT model with time-varying covariates}\label{sec:AFTmodel}
	When AFT models only contain time-fixed covariates, a linear regression model can be assumed for the natural logarithm of the failure times, for $1 \leq i \leq n$, as follows
	\begin{equation}\label{eq:loglinearmod}
		\log T_i = \boldsymbol{x}_i\boldsymbol{\beta} + \varepsilon_i,
	\end{equation}
	where $\boldsymbol{\beta}$ is a $p \times 1$ vector of unknown regression coefficients for the time-fixed covariates. We assume that the $\varepsilon_i$, which represent the error terms, are independently and identically distributed as $\varepsilon$. 
    Adopting hazard functions,
    \eqref{eq:loglinearmod} can be represented as 
	\begin{equation}\label{eq:hazardform}
		\lambda(t|\boldsymbol{x}_i) = \lambda_0(te^{-\boldsymbol{x}_i\boldsymbol{\beta}})e^{-\boldsymbol{x}_i\boldsymbol{\beta}},
 	\end{equation}
	in which $\lambda(t|\boldsymbol{x}_i)$ denotes the hazard function of $T_i$ and $\lambda_0(t)$ represents 
 the hazard function of $T_0 = e^{\varepsilon}$ (called the baseline hazard henceforth). In semiparametric AFT models, the distribution of $\varepsilon$ is unspecified and hence the baseline hazard $\lambda_0(t)$ is unknown. 
 
    Let $\boldsymbol{\gamma}$ be a $q \times 1$ vector of unknown regression coefficients for the time-varying covariates. Following \cite{Cox1987}, 
    the AFT model with time-varying covariates can be written as 
	\begin{equation}\label{eq:hazardformTVC}
        \lambda(t|\boldsymbol{x}_i,\tilde{\boldsymbol{z}}_i(t)) = \lambda_0(\kappa_i(t))e^{-\boldsymbol{x}_i\boldsymbol{\beta}-\boldsymbol{z}_i(t)\boldsymbol{\gamma}},
	\end{equation}
	where 
	\begin{equation}\label{eq:kappaform}
		\kappa_i(t;\boldsymbol{\beta},\boldsymbol{\gamma}) = e^{-\boldsymbol{x}_i\boldsymbol{\beta}}\int^t_0e^{-\boldsymbol{z}_i(s)\boldsymbol{\gamma}}ds.
	\end{equation}
	From the hazard function in \eqref{eq:hazardformTVC}, the cumulative hazard is
	$
		\Lambda(t|\boldsymbol{x}_i,\tilde{\boldsymbol{z}}_i(t)) = \Lambda_0(\kappa_i(t)),
 $
	where $\Lambda_0(\kappa) = \int^\kappa_0\lambda_0(s)ds$ is the cumulative baseline hazard. 
 The survival function can then be expressed as 
$
		S(t|\boldsymbol{x}_i, \tilde{\boldsymbol{z}}_i(t)) = S_0(\kappa_i(t)),
$
	where $S_0(\kappa)= \exp\{-\Lambda_0(\kappa)\}$. To simplify the notation in this article, we will henceforth denote $\lambda(t|\boldsymbol{x}_i,\tilde{\boldsymbol{z}}_i(t))$ and $\Lambda(t|\boldsymbol{x}_i,\tilde{\boldsymbol{z}}_i(t))$ by $\lambda_i(t)$ and $\Lambda_i(t)$ respectively.
	
	Note that the nonparametric baseline hazard, $\lambda_0(\kappa)$, is an infinite-dimensional parameter. Thus, attempting to estimate $\lambda_0(\kappa)$ without any restrictions is an ill-conditioned problem as we only have a finite number of observations. By adopting the method-of-sieves \citep{Geman1982}, we make use of a sum of $m$ basis functions, where $m<<n$, to formulate a smooth approximation to $\lambda_0(\kappa)$ as follows
	\begin{equation}\label{eq:splineSum}
		\lambda_0(\kappa) = \sum_{u=1}^{m}\theta_u\psi_u(\kappa),
	\end{equation}
	where $\theta_u$ is an unknown but fixed coefficient for each basis function $\psi_u(\kappa)$. Each basis function is chosen such that $\psi_u(\kappa) \geq 0$. This ensures that the constraint $\lambda_0(\kappa) \geq 0$ is fulfilled simply by implementing the constraint $\theta_u \geq 0$ for all $u$.
	
	Note that $\kappa$ is dependent on  $\boldsymbol{\beta}$ and $\boldsymbol{\gamma}$.  Since $\boldsymbol{\beta}$ and $\boldsymbol{\gamma}$ are estimated iteratively in our  approach, the boundary points of $\kappa$ constantly shift. To avoid setting inaccurate boundary points for our basis functions, we employ the use of a Gaussian basis function since its boundary points are $-\infty$ and $\infty$ \citep{Ma2023}. 
	
	In accordance with \eqref{eq:splineSum}, a Gaussian basis function is defined by 
	\begin{equation*}
		\psi_u(\kappa) = \frac{\exp\{-(\kappa- \mu_u)^2/(2\sigma_u^2)\}}{\sqrt{2\pi\sigma_u^2}},
	\end{equation*}
	for $ 1 \leq u \leq m$, where $\mu_u$ is defined as the central location, or knot, of a particular basis function. Let $\phi(x)$ and $\Phi(x)$ denote the probability density function and the cumulative distribution function of a standard normal distribution respectively. Then, the above Gaussian basis function can be re-expressed 
 as $\psi_u(\kappa) = \phi((\kappa-\mu_u)/\sigma_u)/\sigma_u$. Similarly, the cumulative basis function can be expressed as $\Psi_u(\kappa) = \int_{0}^{\kappa}\psi_u(s)ds = \Phi((\kappa - \mu_u)/\sigma_u) - \Phi(-\mu_u/\sigma_u)$. Hence, the cumulative baseline hazard may now be expressed as follows
	\begin{equation}\label{eq:splineCumSum}
		\Lambda_0(\kappa) = \sum_{u=1}^{m}\theta_u\Psi_u(\kappa).
	\end{equation}
	
	\section{Maximum penalised likelihood estimation}\label{sec:MPL}
 \subsection{Penalised likelihood}
	Using the hazard formulation of the semiparametric AFT model shown in \eqref{eq:hazardformTVC}, the log-likelihood for the observations $(\boldsymbol{y}_i, \delta_i, \delta_i^L, \delta_i^R, \delta_i^I, \boldsymbol{x}_i, \tilde{\boldsymbol{z}}_i(y_i^*))$, $1 \leq i \leq n$, can be written as
	\small{\begin{equation}\label{eq:oldloglik}
			\begin{aligned}
				\ell(\boldsymbol{\beta},\boldsymbol{\gamma},
				\boldsymbol{\theta})
				&= \sum^n_{i=1}\Big(\delta_i\{\log\lambda_i(y_i) -\Lambda_i(y_i)\} + \delta_i^R\{-\Lambda_i(y_i^L)\} + \delta_i^L\log\{1-S_i(y_i^R)\}\\
				&\qquad + \delta_i^I\log\{S_i(y_i^L)-S_i(y_i^R)\}\Big),
			\end{aligned}
	\end{equation}}
	\normalsize
    where $\kappa_i(t)$ is given as in \eqref{eq:kappaform} and $\lambda_0(\kappa)$ and $\Lambda_0(\kappa)$ are given as in \eqref{eq:splineSum} and \eqref{eq:splineCumSum} respectively. By rewriting the survival functions, the expression in \eqref{eq:oldloglik} can be further evaluated as follows
	\begin{equation}\label{eq:loglik}
		\begin{aligned}
			\ell(\boldsymbol{\beta}, \boldsymbol{\gamma},
			\boldsymbol{\theta})
			&= \sum^n_{i=1}\Big(\delta_i\{\log\lambda_0(\kappa_i(y_i)) - \boldsymbol{x}_i\boldsymbol{\beta} - \boldsymbol{z}_i(y_i)\boldsymbol{\gamma} -\Lambda_0(\kappa_i(y_i))\} - \delta_i^R\Lambda_0(\kappa_i(y_i^L))\\
			&\qquad  + \delta_i^L\log\left[1-\exp\{-\Lambda_0(\kappa_i(y_i^R))\}\right]\\
			&\qquad + \delta_i^I\log\left[\exp\{-\Lambda_0(\kappa_i(y_i^L))\}-\exp\{-\Lambda_0(\kappa_i(y_i^R))\}\right]\Big).
		\end{aligned}
	\end{equation}
	
	To penalise the log-likelihood in \eqref{eq:loglik}, a roughness penalty function is imposed. This penalty not only imposes smoothness on the $\lambda_0(\kappa)$ estimate, but it also reduces the reliance of the estimates on the number of basis functions and the locations of their knots. Hence, greater flexibility and stability in the estimation of the baseline hazard is achieved. In addition to the use of Gaussian basis functions, the penalty also enforces an extra level of smoothness on the approximation to the baseline hazard function. 
	
	Let $\boldsymbol{\theta}$ be the $m \times 1$ vector of unknown but fixed coefficients for the basis functions. The roughness penalty \citep{Eubank1999} that we will use in our approach is given by
	\begin{equation}\label{eq:penalty}
		\int^{d_2}_{d_1}\{\lambda_0^{''}(s)\}^2ds = \boldsymbol{\theta}^T\boldsymbol{R}\boldsymbol{\theta},
	\end{equation}
	where $d_1$ and $d_2$, respectively, denote the minimum and maximum of all $\kappa_i(y_i)$ and each $(u,v)$-th element of matrix $\boldsymbol{R}$ is calculated as $\int^{d_2}_{d_1}\psi_u^{''}(s)\psi_v^{''}(s)ds$. 
	
	Combining \eqref{eq:loglik} and \eqref{eq:penalty}, the penalised log-likelihood can be expressed as
	\begin{equation*}
		P(\boldsymbol{\beta}, \boldsymbol{\gamma}, \boldsymbol{\theta}) = \ell(\boldsymbol{\beta}, \boldsymbol{\gamma}, \boldsymbol{\theta}) - h\boldsymbol{\theta}^T\boldsymbol{R}\boldsymbol{\theta},
	\end{equation*}
	where $ h \geq 0$ is the smoothing parameter and $\boldsymbol{R}$ is assumed to be fixed after a predetermined number of initial iterations. Thus, the maximum  likelihood estimates of $(\boldsymbol{\beta}, \boldsymbol{\gamma}, \boldsymbol{\theta})$ can be obtained by solving the following constrained optimisation problem
	\begin{equation}\label{eq:MPLestimates}
		(\hat{\boldsymbol{\beta}}, \hat{\boldsymbol{\gamma}}, \hat{\boldsymbol{\theta}}) = \underset{\boldsymbol{\beta}, \boldsymbol{\gamma}, \boldsymbol{\theta}}{\text{argmax}} P(\boldsymbol{\beta}, \boldsymbol{\gamma}, \boldsymbol{\theta}),
	\end{equation}
	where $\boldsymbol{\theta} \geq 0$, with the inequality interpreted elementwise.
	
	\subsection{Computation of solutions}
	For the constrained solution to \eqref{eq:MPLestimates}, the Karush-Kuhn-Tucker (KKT) necessary conditions are:
	\begin{equation*}
		\begin{aligned}
			&\frac{\partial P}{\partial \beta_j} = 0, \quad j = 1 \dots p; \\
			&\frac{\partial P}{\partial \gamma_r} = 0, \quad r = 1 \dots q;\\
			&\frac{\partial P}{\partial \theta_u} = 0\text{ if }\theta_u > 0\text{ and } \frac{\partial P}{\partial \theta_u} < 0\text{ if }\theta_u  = 0, \quad u = 1\dots m,
		\end{aligned}
	\end{equation*}
    where the case in which $\theta_u = 0$ with a negative gradient signifies an active constraint.
 
	To solve these equations, following the approach of \cite{Ma2023}, we utilise an algorithm that alternates between pseudo-Newton methods and the MI algorithm within each iteration to update the estimates of $\boldsymbol{\beta}$, $\boldsymbol{\gamma}$, and $\boldsymbol{\theta}$. Specifically, for the pseudo-Newton methods, negative definite terms are extracted from the Hessians to construct pseudo-Hessians. These pseudo-Hessians are then used in the pseudo-Newton methods to update $\boldsymbol{\beta}$ and $\boldsymbol{\gamma}$, ensuring that the search direction is uphill. Subsequently, the MI algorithm is applied to update each $\theta_u$, with $\boldsymbol{\beta}$ and $\boldsymbol{\gamma}$ fixed at their current values. The MI algorithm ensures that all $\theta_u$ remain non-negative. This process is repeated for multiple iterations until convergence is achieved. The procedure is described in more detail below. Note that the analytic expressions of the gradient vectors $\nabla_{\boldsymbol{\beta}}P(\boldsymbol{\beta}, \boldsymbol{\gamma}, \boldsymbol{\theta})$, $\nabla_{\boldsymbol{\gamma}}P(\boldsymbol{\beta}, \boldsymbol{\gamma}, \boldsymbol{\theta})$, and $\nabla_{\boldsymbol{\theta}}P(\boldsymbol{\beta}, \boldsymbol{\gamma}, \boldsymbol{\theta})$, as well as the pseudo-Hessian matrices $H^{-}_{\boldsymbol{\beta}}P(\boldsymbol{\beta}, \boldsymbol{\gamma}, \boldsymbol{\theta})$ and $H^{-}_{\boldsymbol{\gamma}}P(\boldsymbol{\beta}, \boldsymbol{\gamma}, \boldsymbol{\theta})$ are provided in the supplementary material.
	
	To update from $\boldsymbol{\beta}^{(k)}$ to $\boldsymbol{\beta}^{(k+1)}$, we use the following pseudo-Newton method:
	\begin{equation*}
		\boldsymbol{\beta}^{(k+1)} = \boldsymbol{\beta}^{(k)} + a_{\boldsymbol{\beta}}^{(k)}\left\{H^{-}_{\boldsymbol{\beta}}P(\boldsymbol{\beta}^{(k)}, \boldsymbol{\gamma}^{(k)}, \boldsymbol{\theta}^{(k)})\right\}^{-1}\nabla_{\boldsymbol{\beta}}P(\boldsymbol{\beta}^{(k)}, \boldsymbol{\gamma}^{(k)}, \boldsymbol{\theta}^{(k)}),
	\end{equation*}
	where $a_{\boldsymbol{\beta}}^{(k)}$ denotes the stepsize used to ensure that $P(\boldsymbol{\beta}^{(k)}, \boldsymbol{\gamma}^{(k)}, \boldsymbol{\theta}^{(k)}) \leq P(\boldsymbol{\beta}^{(k+1)}, \boldsymbol{\gamma}^{(k)}, \boldsymbol{\theta}^{(k)})$. 
    
    Similarly, to update from $\boldsymbol{\gamma}^{(k)}$ to $\boldsymbol{\gamma}^{(k+1)}$, the following pseudo-Newton method is used:
	\begin{equation*}
		\boldsymbol{\gamma}^{(k+1)} = \boldsymbol{\gamma}^{(k)} + a_{\boldsymbol{\gamma}}^{(k)}\left\{H^{-}_{\boldsymbol{\gamma}}P(\boldsymbol{\beta}^{(k+1)}, \boldsymbol{\gamma}^{(k)}, \boldsymbol{\theta}^{(k)})\right\}^{-1}\nabla_{\boldsymbol{\gamma}}P(\boldsymbol{\beta}^{(k+1)}, \boldsymbol{\gamma}^{(k)}, \boldsymbol{\theta}^{(k)}),
	\end{equation*}
	where $a_{\boldsymbol{\gamma}}^{(k)}$ is the stepsize used to ensure that $P(\boldsymbol{\beta}^{(k+1)}, \boldsymbol{\gamma}^{(k)}, \boldsymbol{\theta}^{(k)}) \leq P(\boldsymbol{\beta}^{(k+1)}, \boldsymbol{\gamma}^{(k+1)}, \boldsymbol{\theta}^{(k)})$. 
	
	After updating the estimates of $\boldsymbol{\beta}$ and $\boldsymbol{\gamma}$, the MI algorithm is used to update $\boldsymbol{\theta}$ while constraining $\boldsymbol{\theta}$ to be non-negative. Thus, to update from $\boldsymbol{\theta}^{(k)}$ to $\boldsymbol{\theta}^{(k+1)}$, we use the following MI scheme:
		\begin{equation}\label{eq:MIscheme}
		\boldsymbol{\theta}^{(k+1)} = \boldsymbol{\theta}^{(k)} + a_{\boldsymbol{\theta}}^{(k)}S_{\boldsymbol{\theta}}P(\boldsymbol{\beta}^{(k+1)}, \boldsymbol{\gamma}^{(k+1)}, \boldsymbol{\theta}^{(k)})
    \nabla_{\boldsymbol{\theta}}P(\boldsymbol{\beta}^{(k+1)}, \boldsymbol{\gamma}^{(k+1)}, \boldsymbol{\theta}^{(k)}),
	\end{equation}
	where $S_{\boldsymbol{\theta}}$ is a diagonal matrix with elements given in Section 1.4.3 of the supplementary material. Here, $a_{\boldsymbol{\theta}}^{(k)}$ is the stepsize that ensures $P(\boldsymbol{\beta}^{(k+1)}, \boldsymbol{\gamma}^{(k+1)}, \boldsymbol{\theta}^{(k)}) \leq P(\boldsymbol{\beta}^{(k+1)}, \boldsymbol{\gamma}^{(k+1)}, \boldsymbol{\theta}^{(k+1)})$ in iteration $k$. Note that the algorithm guarantees that $\boldsymbol{\theta}^{(k+1)} \geq 0$ whenever $\boldsymbol{\theta}^{(k)} \geq 0$.
	
	\subsection{Automatic smoothing parameter selection}
	A crucial part of our penalised likelihood approach is the inclusion of an automatic selection process for the smoothing parameter. Traditional methods may require manual tuning of smoothing parameters, which introduces subjectivity into the estimation process. Automatic selection methods help in avoiding this subjectivity by providing an objective and data-driven approach to choose the smoothing parameter. In addition, the choice of smoothing parameter directly impacts the performance of penalised likelihood methods. Selecting an optimal smoothing parameter helps achieve a balance between fitting the data well and avoiding overfitting.
	
	We adopt a marginal likelihood-based method to estimate the smoothing parameter $h$. As noted by \cite{Cai2003}, \cite{Ma2021} and others, the penalty function can be represented as the log-prior density of $\boldsymbol{\theta}$ where $\boldsymbol{\theta} \sim N (\boldsymbol{0}_{m \times 1}, \sigma_h^2\boldsymbol{R}^{-1})$ with the smoothing parameter introduced through $\sigma_h^2 = 1/{2h}$. The log-posterior can then be expressed as
	\begin{equation}\label{eq:logPosterior}
		\ell_P(\boldsymbol{\beta}, \boldsymbol{\gamma}, \boldsymbol{\theta}, \sigma_h^2) \approx -\frac{m}{2}\log(\sigma^2_h) + 	\ell(\boldsymbol{\beta}, \boldsymbol{\gamma}, \boldsymbol{\theta}) - \frac{1}{2\sigma^2_h}\boldsymbol{\theta}^T\boldsymbol{R}\boldsymbol{\theta}.
	\end{equation}
    From \eqref{eq:logPosterior}, the marginal likelihood of $\sigma^2_h$  can be obtained by integrating out $\boldsymbol{\beta}, \boldsymbol{\gamma}$ and $\boldsymbol{\theta}$ from $\exp\{	\ell_P(\boldsymbol{\beta}, \boldsymbol{\gamma}, \boldsymbol{\theta}, \sigma_h^2)\}$, the posterior density. This integral, however, is intractable. One way to circumvent this issue is by making use of a Laplace approximation to the high-dimensional integral and obtaining an approximate marginal likelihood of $\sigma_h^2$ instead. Taking the logarithm of this approximation gives us
	\begin{equation}\label{eq:marginalLik}
		\ell_m(\sigma^2_h) \approx \frac{d}{2}\ln(2\pi) -\frac{m}{2}\log(\sigma^2_h) + \ell(\hat{\boldsymbol{\beta}}, \hat{\boldsymbol{\gamma}}, \hat{\boldsymbol{\theta}}) - \frac{1}{2\sigma^2_h}\hat{\boldsymbol{\theta}}^T\boldsymbol{R}\hat{\boldsymbol{\theta}} - \frac{1}{2}\log|\hat{\boldsymbol{F}} + \boldsymbol{Q}(\sigma^2_h)|,
	\end{equation}
	where $\hat{\boldsymbol{F}}$ is the negative Hessian of the log-likelihood $\ell(\boldsymbol{\beta}, \boldsymbol{\gamma}, \boldsymbol{\theta})$ evaluated at $\hat{\boldsymbol{\beta}}, \hat{\boldsymbol{\gamma}}$ and $\hat{\boldsymbol{\theta}}$ and
	\begin{equation*}
		\boldsymbol{Q}(\sigma^2_h) = \begin{pmatrix}
			0 & 0 & 0\\
			0 & 0 & 0\\
			0 & 0 & \frac{1}{\sigma^2_h}\boldsymbol{R}
		\end{pmatrix}.
	\end{equation*}
	Since $\partial\ell_m(\sigma^2_h)/\partial\sigma^2_h$ is non-linear in $\sigma^2_h$, it demands an optimisation procedure to obtain the estimate to $\sigma^2_h$. 
 However, notice that the solution maximising $\ell_m(\sigma^2_h)$ satisfies
	\begin{equation}\label{eq:sigmaUpdate}
		\hat{\sigma}^2_h = \frac{\hat{\boldsymbol{\theta}}^T\boldsymbol{R}\hat{\boldsymbol{\theta}}}{m - \nu},
	\end{equation}
	where
	$
		\nu = \text{tr}\{(\hat{\boldsymbol{F}} + \boldsymbol{Q}(\hat{\sigma}^2_h))^{-1}\boldsymbol{Q}(\hat{\sigma}^2_h)\},
$
 which can be interpreted as the model degrees of freedom. Hence, we can simply update $\sigma^2_h$ using this formula where the $\sigma^2_h$ on the right-hand side is replaced by its most current value. 
	
	Since updating the MPL estimates of $\boldsymbol{\beta}, \boldsymbol{\gamma}$ and $\boldsymbol{\theta}$ depends on the estimate of $\sigma^2_h$, we use an alternating scheme to update all of the parameters. Firstly, with the estimate of $\sigma^2_h$ fixed, we obtain the estimates of $\boldsymbol{\beta}, \boldsymbol{\gamma}$ and $\boldsymbol{\theta}$ by maximising the penalised log-likelihood. Then, once the values of $\hat{\boldsymbol{\beta}}, \hat{\boldsymbol{\gamma}}$ and $\hat{\boldsymbol{\theta}}$ are obtained, the estimate of $\hat{\sigma}^2_h$ is updated by \eqref{eq:sigmaUpdate}. This process repeats itself until $\nu$ is stabilised, meaning 
 the fluctuation of $\nu$ is bounded within a pre-determined tolerance level.
	
	\subsection{Asymptotic properties}\label{sec:asymptotics}
	By following the approach of \cite{Ma2021} and \cite{Ma2024}, asymptotic results are readily available for the MPL estimates of $\boldsymbol{\beta}, \boldsymbol{\gamma}$ and $\boldsymbol{\theta}$. Specifically, asymptotic consistency and asymptotic normality are obtained under the conditions listed in \citet[Section 4.4]{Ma2024}. Hence, we will omit the technical details here.
	
	Let $\boldsymbol{\eta}_0 = (\boldsymbol{\theta}_0^T, \boldsymbol{\beta}_0^T, \boldsymbol{\gamma}_0^T)^T$, which denotes the $(m+p+q) \times 1$ vector of true parameter values. In order to facilitate inferences on $\boldsymbol{\beta}, \boldsymbol{\gamma}$ and the baseline hazard, $\lambda_0(t)$, we make use of the asymptotic normality of $\boldsymbol{\eta}$. 
 
    However,  as is often encountered in practice in the MPL approach for survival analysis, some of the $\boldsymbol{\theta} \geq 0$ constraints can be active. This occurs especially when there are more knots than necessary. In such instances, the penalty function will suppress unnecessary $\theta_u$ values to zero. If the effects of these active constraints are not accounted for in the asymptotic normality results, one may encounter unwanted issues, such as obtaining negative asymptotic variances for the estimators of the model parameters. To avoid this, the general result of \cite{Moore2008} and results in \cite{Ma2021} have been adopted to derive asymptotic normality results for the constrained MPL estimators proposed in this paper. The result involves a matrix $\boldsymbol{U}$ to accommodate the active constraints. To define $\boldsymbol{U}$, without loss of generality, assume that the first $\tilde{m}$ of the $\boldsymbol{\theta} \geq 0$ constraints are active. Then $ \boldsymbol{U} = [\boldsymbol{0}_{(m - \tilde{m} +p + q) \times \tilde{m}}, \boldsymbol{I}_{(m - \tilde{m} +p + q) \times (m - \tilde{m} +p + q)}]$, which satisfies $\boldsymbol{U}^T\boldsymbol{U} = \boldsymbol{I}_{(m - \tilde{m} +p + q) \times (m - \tilde{m} +p + q)}$.
		
	As stated in \cite{Ma2024} (see also \cite{Ma2023}),  assuming that the required assumptions are met, the distribution of $\hat{\boldsymbol{\eta}}$ is approximately multivariate normal when $n$ is large. In addition, the covariance matrix can be calculated using
	\begin{equation}\label{eq:asyVar}
		\text{var}(\hat{\boldsymbol{\eta}}) = -\boldsymbol{B}(\hat{\boldsymbol{\eta}})^{-1}\frac{\partial^2\ell(\hat{\boldsymbol{\eta}})}{\partial\boldsymbol{\eta}\partial\boldsymbol{\eta}^T}\boldsymbol{B}(\hat{\boldsymbol{\eta}})^{-1}
	\end{equation}
	where
	\begin{equation*}
		\boldsymbol{B}(\hat{\boldsymbol{\eta}})^{-1} = \boldsymbol{U}\left(\boldsymbol{U}^T\left(-\frac{\partial^2\ell(\hat{\boldsymbol{\eta}})}{\partial\boldsymbol{\eta}\partial\boldsymbol{\eta}^T} +h\boldsymbol{R}\right)\boldsymbol{U}\right)^{-1}\boldsymbol{U}^T.
	\end{equation*}
        Here, the $\boldsymbol{U}$ matrix aids in deleting the rows and columns associated with the active constraints in $-\left(\partial^2\ell(\hat{\boldsymbol{\eta}})/\partial\boldsymbol{\eta}\partial\boldsymbol{\eta}^T\right) +h\boldsymbol{R}$. Then, $\boldsymbol{B}(\hat{\boldsymbol{\eta}})^{-1}$ is obtained by padding the inverse of $\boldsymbol{U}^T\left\{-\left(\partial^2\ell(\hat{\boldsymbol{\eta}})/\partial\boldsymbol{\eta}\partial\boldsymbol{\eta}^T\right) +h\boldsymbol{R}\right\}\boldsymbol{U}$ with zeros in the deleted rows and columns.
	
	In practical terms, identifying active constraints involves assessing both the value of $\theta_u$ and its associated gradient. Following the convergence of the Newton-MI algorithm, certain $\theta_u$ values may precisely reach zero with negative gradients, clearly designating them as active constraints. However, there may be $\theta_u$ values that are very close to, but not exactly at, zero. In such cases, a negative gradient indicates that they are also subject to an active constraint. Therefore, in practical terms, active constraints are determined where, for a given $u$, $\theta_u < 10^{-2}$ and the corresponding gradient is less than $-\epsilon$, where $\epsilon$ represents a positive threshold value such as $10^{-2}$.
	
	\section{A simulation study}\label{sec:simulation}
	A simulation study was conducted to evaluate the methodology proposed in this paper. To the best of our knowledge, there are currently no other approaches or publicly available \texttt{R} packages for fitting semiparametric AFT models with time-varying covariates under partly interval censoring. Therefore, the simulation study will solely assess the accuracy of the estimates obtained using our approach. We first explain the data generation mechanism.
	
	For this particular simulation study, there are two time-fixed covariates included, with the values  of $x_{i1}$ and $x_{i2}$ generated according to a Bernoulli distribution, $\text{Bernoulli}(0.5)$, and a uniform distribution, $\text{Unif}[0, 3]$, respectively. We also assumed a single time-varying covariate, $z_i(t)$, which is a discrete function of $t$ containing a single change-point, defined as
	\begin{equation}\label{eq:tvceg}
		z_i(t) = \begin{cases}
			0 & \text{if } 0 < t < \tau_i;\\
			1 & \text{if } t \geq \tau_i.
		\end{cases}
	\end{equation}
	Here, $\tau_i > 0$ represents the time at which individual $i$ receives a particular treatment.  We generated $\tau_i$ from a uniform distribution: $\tau_i \sim \text{Unif}[0,1]$. The hazard for $T_0 = e^\epsilon$ was chosen to follow a Weibull hazard given by $\lambda_0(t) = 3t^2$, and 
 the corresponding baseline survival function is $S_0(t) = \exp(-t^3)$. In accordance with \eqref{eq:tvceg}, the survival function for an individual $i$ can subsequently be written as 
	\begin{equation*}\label{eq:survival}
		S_i(t) = \begin{cases}
			 \exp\{-(te^{-\boldsymbol{x}_i\boldsymbol{\beta}})^3\} & \text{if } 0 < t < \tau_i;\\
			 \exp[-\{\tau_ie^{-\boldsymbol{x}_i\boldsymbol{\beta}} +(t- \tau_i)e^{-\boldsymbol{x}_i\boldsymbol{\beta} - \gamma}\}^3]& \text{if } t \geq \tau_i.
		\end{cases}
	\end{equation*}
 We then use the inverse transform sampling approach to simulate the exact failure times. For $i = 1\dots, n$, we first generate a standard uniform random number, $u_i \sim \text{Unif}[0,1]$. Then, the event time of individual $i$ can be simulated as follows: if $e^{\boldsymbol{x}_i\boldsymbol{\beta}}\sqrt[3]{-\log u_i} < \tau_i$ then $t_i = e^{\boldsymbol{x}_i\boldsymbol{\beta}}\sqrt[3]{-\log u_i}$; otherwise, $t_i = \tau_i + e^{\gamma}(e^{\boldsymbol{x}_i\boldsymbol{\beta}}\sqrt[3]{-\log u_i} - \tau_i)$. After simulating the event time $t_i$, 
 its corresponding interval-censored times were 
 generated 
 according to the following specifications (see  \cite{Webb2023} or \cite{Cai2003}):
	\begin{equation*}
		y_i^L = t_i^{I(U_i^E < \pi^E)}(\alpha_LU_i^L)^{I(\pi^E \leq U_i^E, \alpha_LU_i^L \leq t_i \leq \alpha_RU_i^R)}(\alpha_RU_i^R)^{I(\pi^E \leq U_i^E, \alpha_RU_i^R < t_i)}0^{I(\pi^E \leq U_i^E, t_i <\alpha_LU_i^L)}
	\end{equation*}
	and
	\begin{equation*}
		y_i^R = t_i^{I(U_i^E < \pi^E)}(\alpha_LU_i^L)^{I(\pi^E \leq U_i^E, t_i < \alpha_LU_i^L)}(\alpha_RU_i^R)^{I(\pi^E \leq U_i^E, \alpha_LU_i^L \leq t_i \leq \alpha_RU_i^R)}\infty^{I(\pi^E \leq U_i^E, \alpha_RU_i^R < t_i)},
	\end{equation*}
	where $\pi^E$ denotes the event proportion and $U_i^E, U_i^L$ and $U_i^R$ denote independent standard uniform random variables for generating the event times and interval censoring times. In addition, $\alpha_L$ and $\alpha_R$ denote scalar quantities that control the width of the censoring intervals and hence, can affect the proportions of events that are left, right or interval-censored. 
	
	In this simulation, we set the true values of $\beta_1$ and $\beta_2$ to $1$ and $-1$ respectively, and the true value of $\gamma$ to $-0.1$. Two sample sizes, $n = 100$ and $n = 1000$, were chosen to represent both small and large samples. Additionally, we considered two levels of event proportions: $\pi^E = 0.3$ and $\pi^E = 0.7$. Let $\pi^L$, $\pi^I$, and $\pi^R$ represent the proportions of left, interval, and right censoring respectively. When $\pi^E = 0.3$, the average proportions for different types of censoring
 were $\pi^L = 0.17$, $\pi^I = 0.20$, and $\pi^R = 0.33$. Conversely, when $\pi^E = 0.7$, 
 we had $\pi^L = 0.08$, $\pi^I = 0.14$, and $\pi^R = 0.08$ on average. 
 In the fitted models, the baseline hazard was approximated 
 using Gaussian basis functions as described previously. The number of basis functions used, $m$, was set to the cubic root of the sample size $n$. The 
 knots (i.e. means of the Gaussian functions) were chosen using quantiles. 
	The Newton-MI algorithm 
 discussed in Section \ref{sec:MPL} were used to estimate the parameters, where 
 we fixed 
 the boundary values of the $\kappa_i(t_i)$s' after 100 iterations. 
 
    We present the results of our simulation study in Table \ref{table:tab2} and compare them with those obtained by fitting an AFT model that only accommodates right-censored observations along with the time-varying covariates, as discussed in \cite{Ma2023}, where 
    the censoring intervals for left and interval-censored observations were replaced using midpoint imputation. 
    
    In the following Tables \ref{table:tab2} and \ref{table:tab3}, we report values of the bias, Monte Carlo standard deviation (MCSD), average asymptotic standard deviation (AASD) and the coverage probabilities (CP) calculated from 
    $95\%$ confidence intervals using both the MCSDs and AASDs for both approaches. The MCSDs are compared with the associated AASDs to assess the accuracy of the asymptotic variance formula given in \eqref{eq:asyVar}. 
	
  \begin{table}[ht]
  \caption{Results obtained for $\hat{\beta}_1$, $\hat{\beta}_2$ and $\hat{\gamma}$ for $n = 100, m = 5$ and varying exact event proportions. These results compare the two approaches that accommodate interval and right censoring respectively, based on 200 repetitions.}
		\label{table:tab2}
		\centering
		\begin{tabular}{lllllll}
			\hline
			& \multicolumn{3}{c}{Interval-censoring}                                                            & \multicolumn{3}{c}{Midpoint imputation}                                                            \\ \hline
			& \multicolumn{3}{c}{$\pi^E = 0.3$}                                                          & \multicolumn{3}{c}{$\pi^E = 0.3$}                                                         \\ \cline{2-7}
			& \multicolumn{1}{c}{$\hat{\beta}_1$} & \multicolumn{1}{c}{$\hat{\beta}_2$} & \multicolumn{1}{c}{$\hat{\gamma}$} & \multicolumn{1}{c}{$\hat{\beta}_1$} & \multicolumn{1}{c}{$\hat{\beta}_2$} & \multicolumn{1}{c}{$\hat{\gamma}$} \\ \cline{2-7} 
			Bias      & -0.0037                             & -0.0545                             & -0.1101                            & -0.4755                             & 0.4455                              & 0.2451                             \\ \hline
MCSD      & 0.1088                              & 0.0792                              & 0.2172                             & 0.1115                              & 0.0802                              & 0.2163                             \\ \hline
AASD      & 0.1175                              & 0.0612                              & 0.1933                             & 0.0896                              & 0.0486                              & 0.1581                             \\ \hline
CP (MCSD) & 0.950                               & 0.890                               & 0.925                              & 0.005                               & 0                                   & 0.785                              \\ \hline
CP (AASD) & 0.965                               & 0.805                               & 0.865                              & 0                                   & 0                                   & 0.615                              \\ \hline
			& \multicolumn{3}{c}{$\pi^E = 0.7$}                                                        & \multicolumn{3}{c}{$\pi^E = 0.7$}                                                        \\ \cline{2-7} 
			& \multicolumn{1}{c}{$\hat{\beta}_1$} & \multicolumn{1}{c}{$\hat{\beta}_2$} & \multicolumn{1}{c}{$\hat{\gamma}$} & \multicolumn{1}{c}{$\hat{\beta}_1$} & \multicolumn{1}{c}{$\hat{\beta}_2$} & \multicolumn{1}{c}{$\hat{\gamma}$} \\ \cline{2-7} 
			Bias      & -0.0080                             & -0.0185                             & -0.0576                            & -0.3124                             & 0.2881                              & 0.2744                             \\ \hline
MCSD      & 0.0904                              & 0.0624                              & 0.1798                             & 0.1286                              & 0.1015                              & 0.1957                             \\ \hline
AASD      & 0.0854                              & 0.0450                              & 0.1460                             & 0.0803                              & 0.0452                              & 0.1421                             \\ \hline
CP (MCSD) & 0.955                               & 0.935                               & 0.945                              & 0.280                               & 0.165                               & 0.720                              \\ \hline
CP (AASD) & 0.945                               & 0.840                               & 0.890                              & 0.105                               & 0.055                               & 0.515                              \\ \hline
 \end{tabular} 
	\end{table}

      \begin{table}[ht]
  \caption{Results obtained for $\hat{\beta}_1$, $\hat{\beta}_2$ and $\hat{\gamma}$ for $n = 1000, m = 10$ and varying exact event proportions. These results compare the two approaches that accommodate interval and right censoring respectively, based on 200 repetitions.}
		\label{table:tab3}
		\centering
		\begin{tabular}{lllllll}
			\hline
			& \multicolumn{3}{c}{Interval-censoring}                                                            & \multicolumn{3}{c}{
   Midpoint imputation}                                                            \\ \hline
			& \multicolumn{3}{c}{$\pi^E = 0.3$}                                                          & \multicolumn{3}{c}{$\pi^E = 0.3$}                                                         \\ \cline{2-7}
			& \multicolumn{1}{c}{$\hat{\beta}_1$} & \multicolumn{1}{c}{$\hat{\beta}_2$} & \multicolumn{1}{c}{$\hat{\gamma}$} & \multicolumn{1}{c}{$\hat{\beta}_1$} & \multicolumn{1}{c}{$\hat{\beta}_2$} & \multicolumn{1}{c}{$\hat{\gamma}$} \\ \cline{2-7} 
			Bias      & 0.0041                              & -0.0109                             & -0.0212                            & -0.4006                             & 0.3858                              & 0.1268                             \\ \hline
MCSD      & 0.0346                              & 0.0202                              & 0.0563                             & 0.0346                              & 0.0246                              & 0.0628                             \\ \hline
AASD      & 0.0335                              & 0.0195                              & 0.0572                             & 0.0334                              & 0.0186                              & 0.0568                             \\ \hline
CP (MCSD) & 0.940                               & 0.925                               & 0.940                              & 0                                   & 0                                   & 0.435                              \\ \hline
CP (AASD) & 0.925                               & 0.920                               & 0.955                              & 0                                   & 0                                   & 0.375                              \\ \hline
 & \multicolumn{3}{c}{$\pi^E = 0.7$}                                                        & \multicolumn{3}{c}{$\pi^E = 0.7$}                                                        \\ \cline{2-7} 
			& \multicolumn{1}{c}{$\hat{\beta}_1$} & \multicolumn{1}{c}{$\hat{\beta}_2$} & \multicolumn{1}{c}{$\hat{\gamma}$} & \multicolumn{1}{c}{$\hat{\beta}_1$} & \multicolumn{1}{c}{$\hat{\beta}_2$} & \multicolumn{1}{c}{$\hat{\gamma}$} \\ \cline{2-7} 
			Bias      & 0.0019                              & 0.0030                              & 0.0056                             & -0.2609                             & 0.2430                              & 0.1837                             \\ \hline
MCSD      & 0.0269                              & 0.0182                              & 0.0473                             & 0.0366                              & 0.0260                              & 0.0769                             \\ \hline
AASD      & 0.0253                              & 0.0143                              & 0.0448                             & 0.0286                              & 0.0160                              & 0.0499                             \\ \hline
CP (MCSD) & 0.950                               & 0.935                               & 0.960                              & 0                                   & 0                                   & 0.345                              \\ \hline
CP (AASD) & 0.925                               & 0.890                               & 0.935                              & 0                                   & 0                                   & 0.095                              \\ \hline
\end{tabular} 
	\end{table}
	
	The results show that the biases are generally small across sample sizes when using the approach that accounts for interval censoring. In contrast, using a model that only accommodates right censoring via midpoint imputation leads to larger biases and inadequate coverage probabilities. Additionally, for our approach, the coverage probabilities based on MCSDs are close to the nominal 95\%, indicating a proper balance between bias and MCSDs.

    We also observe that both MCSDs and AASDs improve significantly with increased sample size for our approach. These values further improve as the proportion of exact event times increases. The AASDs closely match the MCSDs for $\hat{\beta}_1$ and $\hat{\gamma}$ when $n = 100$. However, when $n=100$, some discrepancies are noted between the MCSDs and AASDs for $\hat{\beta}_2$. As the sample size increases, the asymptotic accuracy of the AASDs improves, with these discrepancies disappearing at $n = 1000$ in Table \ref{table:tab3}.
    
	\begin{figure}[ht]
		\centering
		\begin{subfigure}[b]{0.499\textwidth}
			\centering
			\caption{$\pi^E = 0.3, n = 100, m = 5$}
			\includegraphics[width=\textwidth]{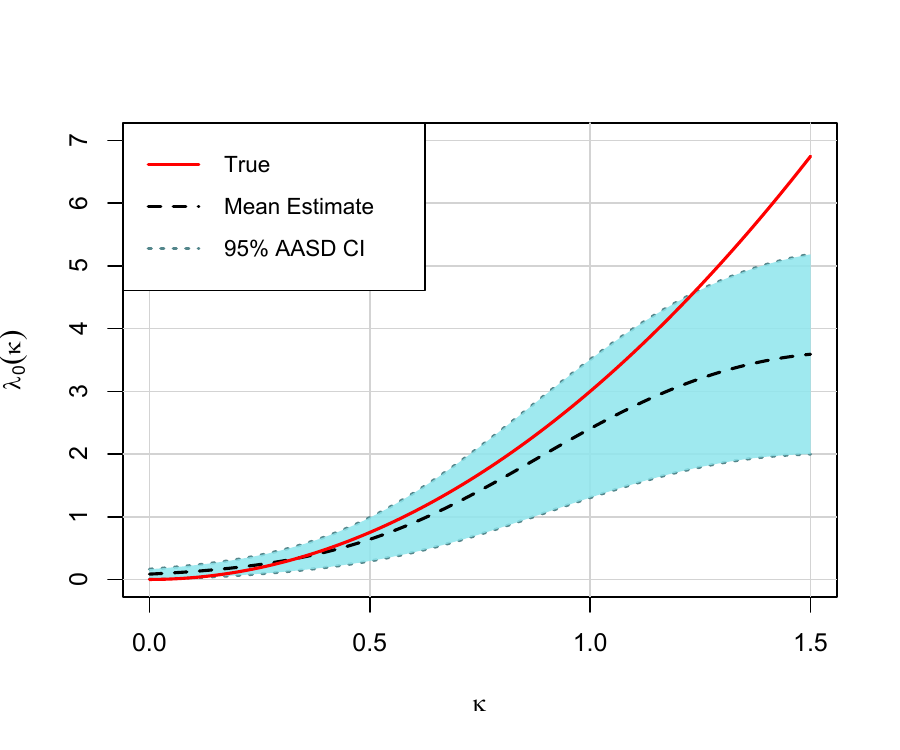}
		\end{subfigure}
		\hfill
		\begin{subfigure}[b]{0.49\textwidth}
			\centering
			\caption{$\pi^E = 0.7, n = 100, m = 5$}
			\includegraphics[width=\textwidth]{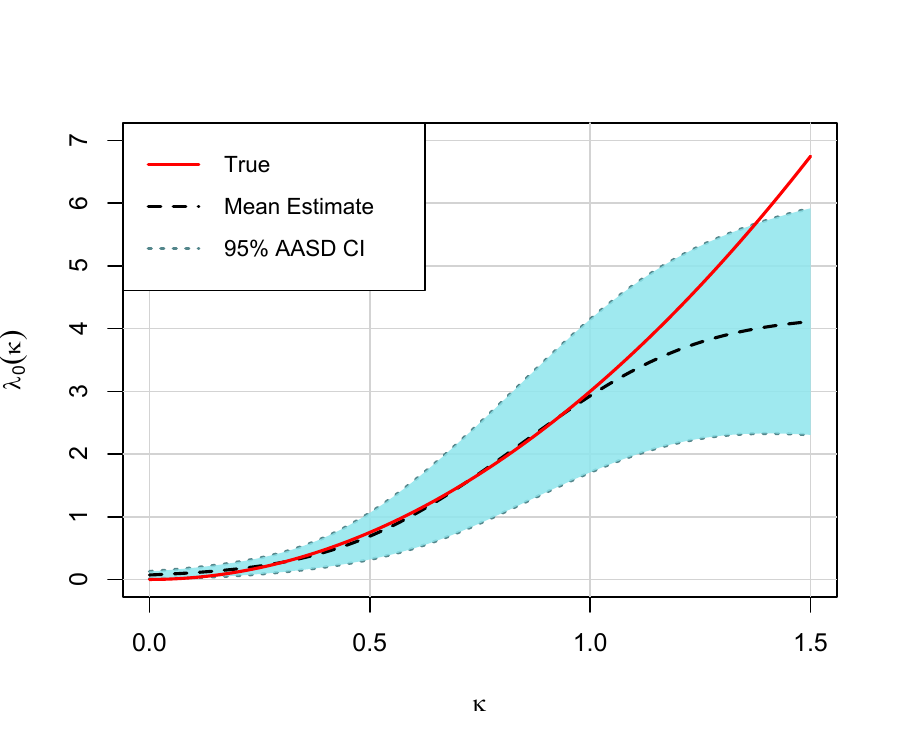}
		\end{subfigure}
		\hfill
		\begin{subfigure}[b]{0.49\textwidth}
			\centering
			\caption{$\pi^E = 0.3, n = 1000, m = 10$}
			\includegraphics[width=\textwidth]{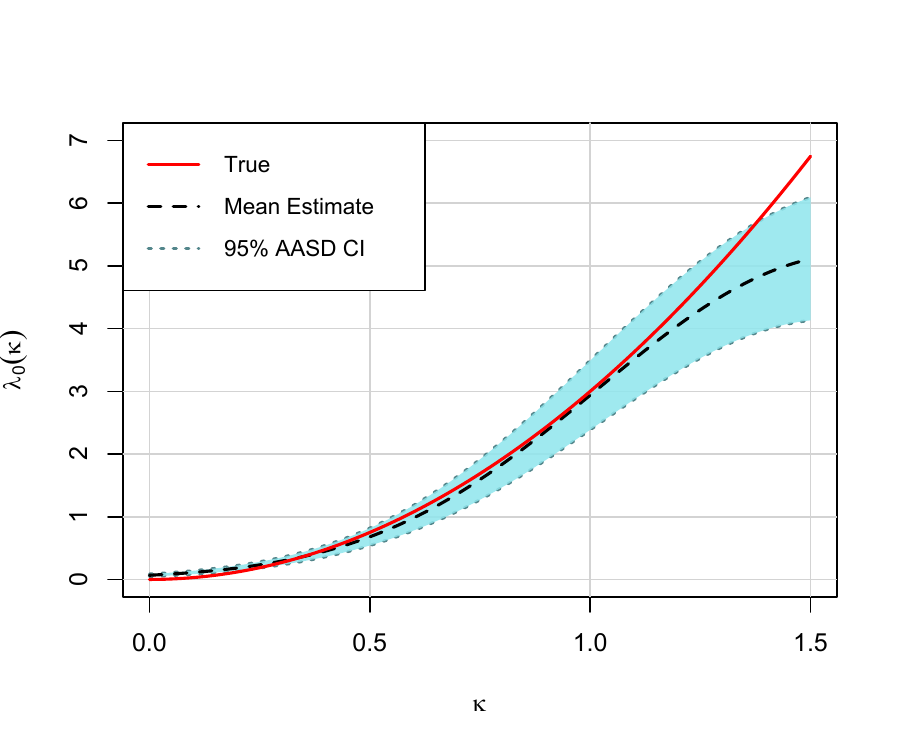}
		\end{subfigure}
		\hfill
		\begin{subfigure}[b]{0.49\textwidth}
			\centering
			\caption{$\pi^E = 0.7, n = 1000, m = 10$}
			\includegraphics[width=\textwidth]{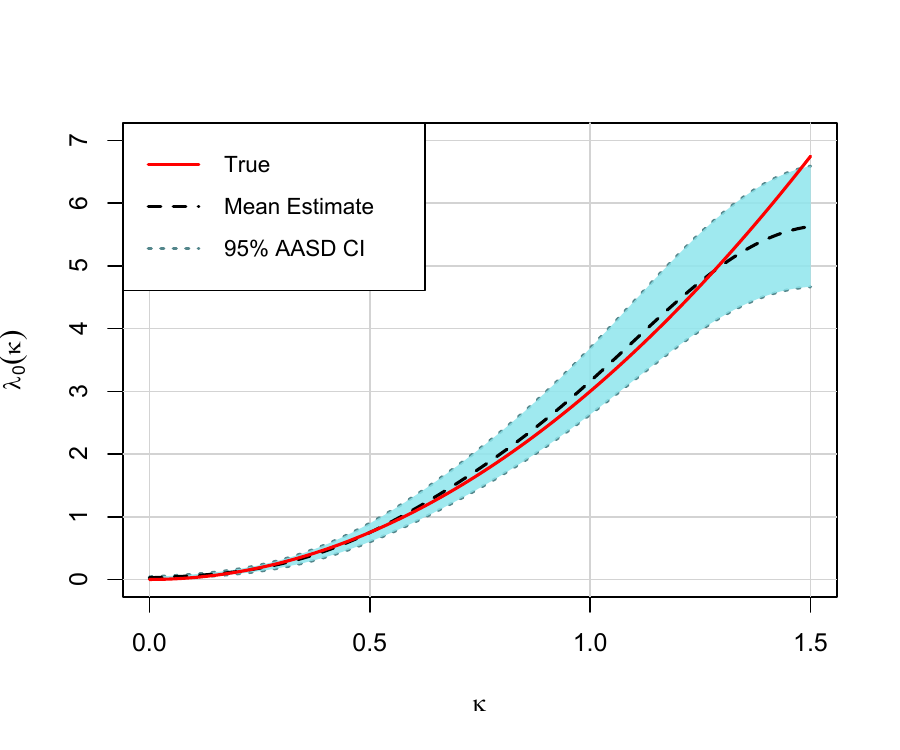}
		\end{subfigure}
		\caption{Figures of the baseline hazard functions on a grid of $\kappa$ values. Each plot includes the true baseline hazard curve (solid red line), the mean estimated baseline hazard curve (dashed black line) and area bound by the 95\% pointwise confidence interval generated using AASD (coloured region).}
		\label{fig:weibullHazPlots}
	\end{figure}

    \begin{figure}[ht]
		\centering
		\begin{subfigure}[b]{0.49\textwidth}
			\centering
			\caption{$\pi^E = 0.3, n = 100, m = 5$}
			\includegraphics[width=\textwidth]{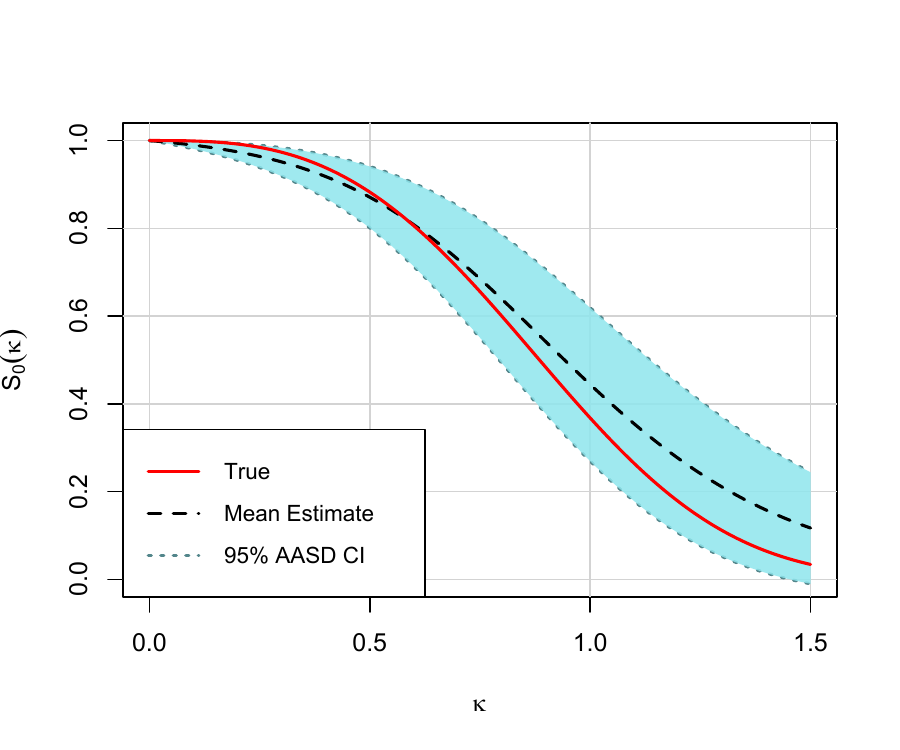}
		\end{subfigure}
		\hfill
		\begin{subfigure}[b]{0.49\textwidth}
			\centering
			\caption{$\pi^E = 0.7, n = 100, m = 5$}
			\includegraphics[width=\textwidth]{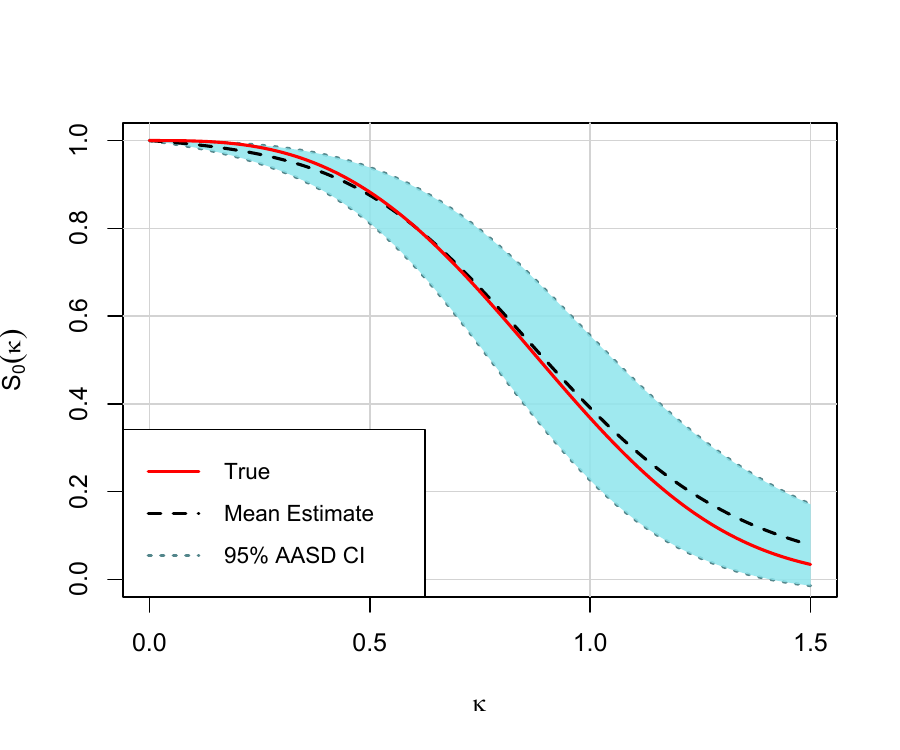}
		\end{subfigure}
		\hfill
		\begin{subfigure}[b]{0.49\textwidth}
			\centering
			\caption{$\pi^E = 0.3, n = 1000, m = 10$}
			\includegraphics[width=\textwidth]{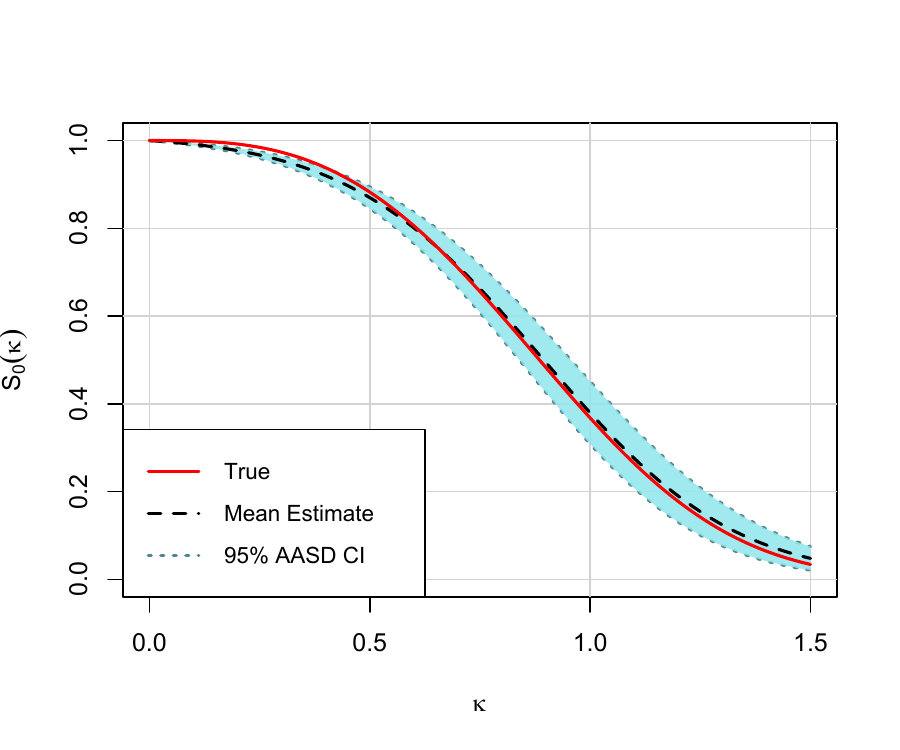}
		\end{subfigure}
		\hfill
		\begin{subfigure}[b]{0.49\textwidth}
			\centering
			\caption{$\pi^E = 0.7, n = 1000, m = 10$}
			\includegraphics[width=\textwidth]{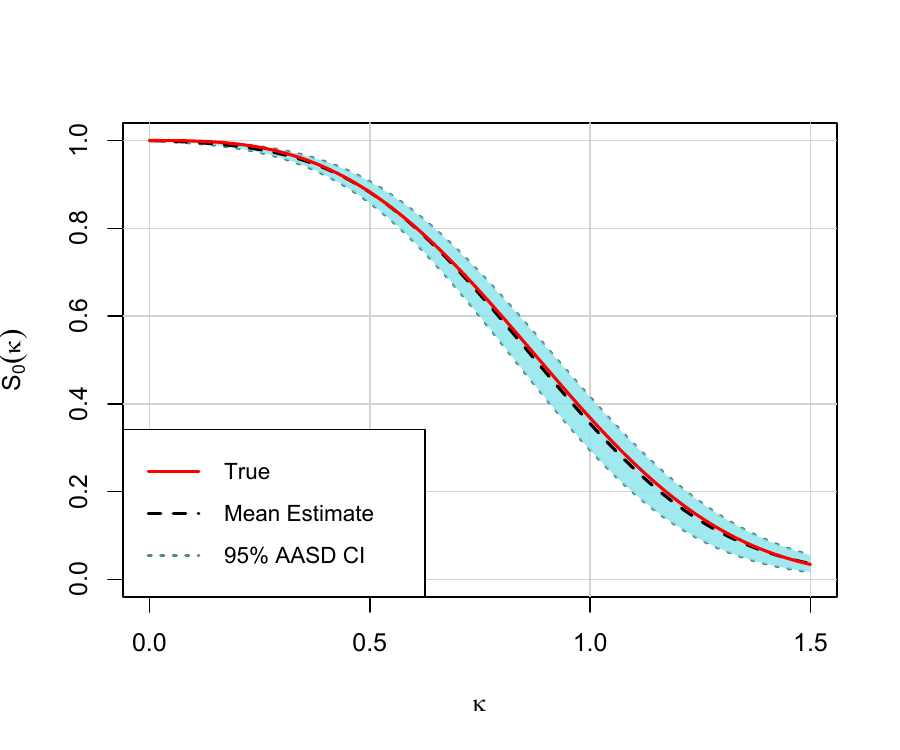}
		\end{subfigure}
		\caption{Figures of the baseline survival functions on a grid of $\kappa$ values. Each plot includes the true baseline survival curve (solid red line), the mean estimated baseline survival curve (dashed black line) and area bound by the 95\% pointwise confidence interval generated using AASD (coloured region).}
		\label{fig:weibullSurvPlots}
	\end{figure}

 Figures \ref{fig:weibullHazPlots} and \ref{fig:weibullSurvPlots} contain plots, based on our approach, illustrating the mean estimated baseline hazard, $\lambda_0(\kappa)$, and the mean estimated baseline survival, $S_0(\kappa)$,  against $\kappa$ for varying proportions of exact event times and sample sizes. The results reveal that the averages of the estimated baseline functions closely align with the true baseline functions especially when $\kappa \leq 1$. An improvement in the accuracy of these estimated baseline functions can be observed over a longer range with an increase in either the proportion of exact failure times or the sample size. In addition, the widths of the pointwise AASD CIs significantly decrease with an increase in sample size, highlighting the improved precision of the estimation. Notably in Figure \ref{fig:weibullHazPlots}, the deviations between the true and mean estimated baseline hazards towards the ends of the curves are attributed to the use of a mixture of Gaussian basis functions in estimating the monotonic Weibull hazard.

\section{Application}\label{sec:application}

\subsection{Main Analysis} \label{subsec:mainAnalysis}
In this section, the WBRTMel trial data, previously analysed in \cite{Hong2019}, is re-analysed using an AFT model adjusted with systemic therapy as a time-varying covariate to evaluate the effectiveness of WBRT treatment directly on the time to intracranial failure. The trial dataset considered includes basic demographic information and medical records for 207 patients (100 patients were randomly assigned to the WBRT arm and 107 patients to the observation arm). The time-to-event is the time to intracranial failure (in years). Of the full cohort, 86 patients (41.5\%) experienced interval-censored failure times, 28 patients (13.5\%) experienced left-censored failure times, and the remaining 93 (45\%) of time-to-event records are right-censored. Notably, there were no exactly observed events in this dataset.

Besides the treatment group (observation or WBRT), the AFT model was adjusted for four additional time-fixed covariates: gender (female or male), presence of extracranial disease at baseline (absent or present), number of melanoma brain metastases (MBM) (1 or 2-3) and age (in years). 

The AFT model also incorporated a single time-varying covariate 
representing the status of systemic therapy. 
During the treatment process, each patient could receive systemic therapy over multiple time periods, where a value of 1 indicated that systemic therapy was active, and 0 indicated it was inactive. Consequently, each patient could have multiple instances of 0's and 1's as time progressed, reflecting changes in their systemic therapy status. 
In this dataset, the number of 0's and 1's per patient ranged from 1 to 44, with a total of 1,187 recorded values. For 12 patients with missing systemic therapy data, the values were replaced with 
``No" in the analysis. 
\begin{table}[t]
\footnotesize
\centering
\caption{Covariates of interest in the WBRTMel dataset}
\label{table_WBRTMel_covariates}
\begin{tabular}{llrr} 
\hline
Covariates & Levels & \# Patients & Percentage \\
\hline
\multirow{2}{*}{Treatment} & Observation (reference group) & 107 & 51.7\% \\
 & WBRT & 100 & 48.3\% \\
 & & & \\
\multirow{2}{*}{Gender} & Female (reference group) & 69 & 33.3\% \\
 & Male  & 138 & 66.7\% \\
 & & & \\
\multirow{2}{*}{Number of MBM} & $=1$ (reference group) & 127 & 61.4\% \\
& $>1$ & 80 & 38.6\% \\
 & & & \\
\multirow{2}{*}{Extracranial disease} & Absent (reference group) & 68 & 32.9\% \\
& Present & 139 & 67.1\% \\
\hline
 & Value & Mean (median) & Standard deviation \\
\hline
Age & (in years) & 61.3 (63.3) & 12.2 \\
\hline
 & Levels & Frequency & \\
\hline
\multirow{2}{*}{Systemic therapy} & No & 1029 & \\
& Yes & 158 & \\ 
\hline

\end{tabular}
\end{table}

A summary of the covariates is shown in Table \ref{table_WBRTMel_covariates}. Two-thirds of the patients were male and the average age was 61.3 years (SD=12.2 years). At the time of randomisation, two-thirds of the patients had extracranial disease and nearly 40\% had more than one MBM. With regards to the time-varying systemic therapy, in addition to the longitudinal records of the total frequency of therapy shown in the table,
44\% of the patients received systemic therapy at some point during the study. 


In order to fit an AFT model to this data set,
we approximated the baseline hazard $\lambda_0(t)$ using 
5 basis functions with their locations determined using a quantile function. Using more than 5 knots resulted in a higher number of active constraints. The optimisation process was halted once the changes in the estimates of the model parameters (all the regression coefficients) were within a tolerance level of $10^{-6}$ and the smoothing parameters had converged.

The results, including the estimated regression coefficients and their corresponding standard errors, are presented in Table \ref{table:coefficients}. Since the estimators for the regression and basis coefficients are asymptotically normal, we also provide the Wald test statistics and p-values in the table. From the table, it is evident that the following covariates are significant: treatment group, number of MBM and systemic therapy.

\begin{table}[ht]
  \caption{Results for the covariates in the WBRTMel dataset using an AFT model}
		\label{table:coefficients}
		\centering
		\begin{tabular}{lrrrr}
			\hline
		& \multicolumn{1}{c}{$\hat{\beta}$} & \multicolumn{1}{c}{se($\hat{\beta}$)} & \multicolumn{1}{c}{Wald statistic} & \multicolumn{1}{c}{P-value}    \\ \hline
			Treatment (WBRT vs observation) & 0.478 & 0.210 & 2.279 & 0.023\\
			Gender (male vs female) & -0.143 & 0.168 & -0.854 & 0.393\\ 
			Number of MBM ($> 1$ vs $= 1$) & -0.498 & 0.168 & -2.960 & 0.003\\
            Extracranial disease (present vs absent) & -0.192 & 0.232 & -0.825 & 0.409\\ 
            Age (in years) & 0.006 & 0.003&1.644&0.100\\ \hline
            & \multicolumn{1}{c}{$\hat{\gamma}$} & \multicolumn{1}{c}{se($\hat{\gamma}$)} & \multicolumn{1}{c}{Wald statistic} & \multicolumn{1}{c}{P-value} \\ \hline
			Systemic therapy (yes vs no) &  0.593 & 0.289 & 2.050 & 0.040 \\ \hline
		\end{tabular} 
	\end{table} 

\begin{table}[ht]
  \caption{Results for the covariates in the WBRTMel dataset using a Cox model}
		\label{table:cox_coefficients}
		\centering
		\begin{tabular}{lrrrr}
			\hline
		& \multicolumn{1}{c}{$\hat{\beta}$} & \multicolumn{1}{c}{se($\hat{\beta}$)} & \multicolumn{1}{c}{Wald statistic} & \multicolumn{1}{c}{P-value}    \\ \hline
			Treatment (WBRT vs observation) & -0.391 & 0.171 & -2.293 & 0.022\\
			Gender (male vs female) & 0.234 & 0.182 & 1.287 & 0.198\\ 
			Number of MBM ($> 1$ vs $= 1$) & 0.226 & 0.171 & 1.323 & 0.198\\
            Extracranial disease (present vs absent) & 0.090 & 0.170 & 0.526 & 0.599\\ 
            Age (in years) & -0.003 & 0.003 &-0.918& 0.359\\ \hline
            & \multicolumn{1}{c}{$\hat{\gamma}$} & \multicolumn{1}{c}{se($\hat{\gamma}$)} & \multicolumn{1}{c}{Wald statistic} & \multicolumn{1}{c}{P-value} \\ \hline
			Systemic therapy (yes vs no) &  -0.353 & 0.001 & -398.595 & $<2e-16$ \\ \hline
		\end{tabular} 
	\end{table}

For comparison's sake, we also fitted a Cox model under  
partly-interval censored data 
with time-varying covariates \citep{Webb2023}, and the results are presented in Table \ref{table:cox_coefficients}. 
As expected, the estimated coefficients exhibit opposite signs between the two models. In addition, both models identify the treatment and systemic therapy covariates as significant. 

Our findings provide new insights into the efficacy of WBRT in treating advanced melanoma patients. These results, obtained from the AFT model adjusted for systemic therapy as a time-varying covariate and other baseline variables such as age, gender, number of MBM and presence of extracranial disease, 
contrast with the 
conclusions 
of \cite{Hong2019}.
Their study, which utilised a univariate logistic regression model and a 
Fine and Gray model without covariate adjustment, reported non-effectiveness of WBRT in treating MBM after local therapy.


\subsection{Dynamic prediction}

Evaluating the efficacy of systemic therapy for a patient is a key application of the model we developed earlier. One specific evaluation can be made through dynamic predictions, which aids clinicians in making informed decisions. In this section, we demonstrate how to calculate the conditional probability that patient $i$ survives beyond $\tau_i + dt$, where $dt$ reflects the change in time,  given that systemic therapy is administered at time $\tau_i$. 

For illustrative purposes, we simplify the form of the time-varying covariate in the WBRTMel dataset as analysed in Section \ref{subsec:mainAnalysis}, $z_i(t)$, to be
	\begin{equation*}\label{eq:tvc}
		z_i(t) = \begin{cases}
			0 & \text{if } 0 < t < \tau_i;\\
			1 & \text{if } \tau_i \leq t < \tau_i + 3/12;\\
			0 & \text{if } t \geq \tau_i + 3/12,
		\end{cases}
	\end{equation*}
where $\tau_i > 0$ is the time when an individual $i$ started receiving systemic therapy for a duration of up to three consecutive months. For comparison, we also compute the same probabilities assuming no systemic therapy is given at $\tau_i$. 

In addition, to evaluate the impact of WBRT, we generate conditional probability curves
 either assuming that the patient received WBRT ($x_{i1} = 1$) or did not receive it ($x_{i1} = 0$). The other covariates are fixed at the following values: Gender = Female ($x_{i2} = 0$), MBM $> 1$ ($x_{i3} = 1$), and Extracranial disease = Present ($x_{i4} = 1$). We use four different values of $\tau_i$ for the calculations: $\tau_i = 0$, $\tau_i = 0.5$, $\tau_i = 1$ and $\tau_i = 1.5$.

The conditional probability we wish to calculate is: 
\begin{equation}
\label{eq:dynamicPred}
    \begin{aligned}
        P(T_i > \tau_i + dt | T_i > \tau_i, 
        \boldsymbol{x}_i, \tilde{\boldsymbol{z}}_i(t), 
        \tau_i) 
        &= \frac{S(\tau_i + dt\,|\,  \boldsymbol{x}_i, \tilde{\boldsymbol{z}}_i(\tau_i+dt))} 
        {S(\tau_i \,|\, \boldsymbol{x}_i, \tilde{\boldsymbol{z}}_i(\tau_i)) },
    \end{aligned}
\end{equation}
with $\boldsymbol{x}_i=(x_{i1}, x_{i2}, x_{i3}, x_{i4})$, and $\tilde{\boldsymbol{z}}_i(t)$ as defined in Section \ref{sec:notation}.

\begin{figure}[ht!]
\centering
\begin{subfigure}[b]{0.49\textwidth}
\centering
\includegraphics[width=\textwidth]{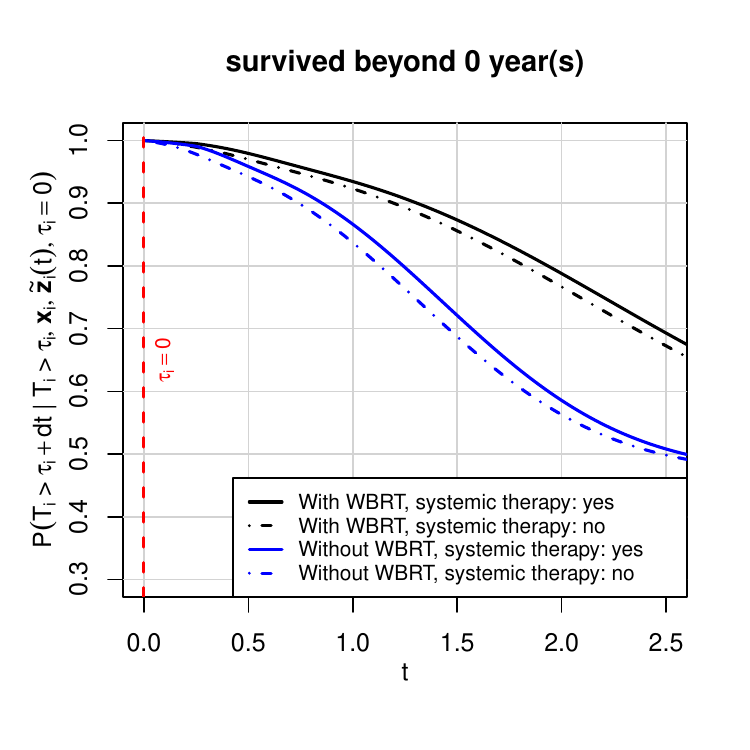}
\end{subfigure}
\hfill
\begin{subfigure}[b]{0.49\textwidth}
\centering
\includegraphics[width=\textwidth]{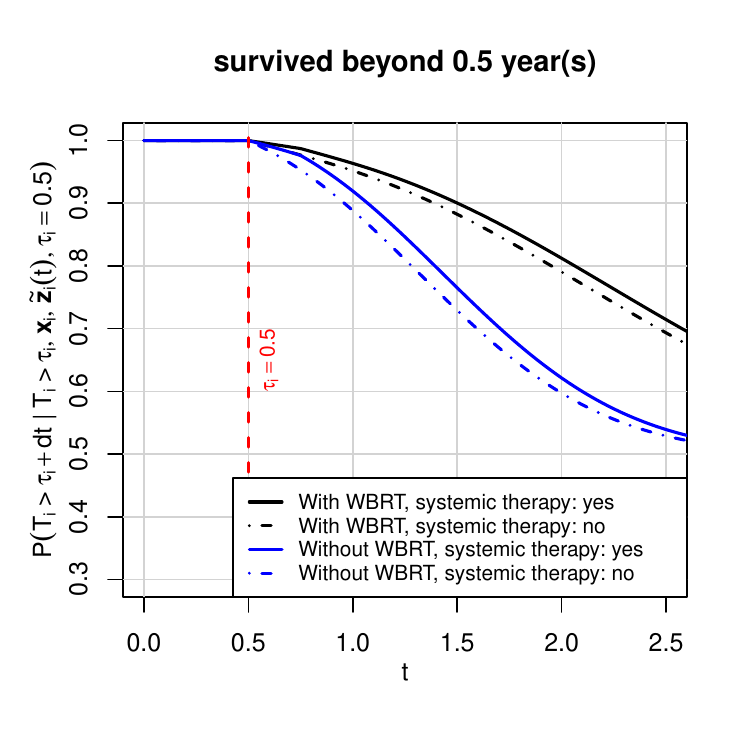}
\end{subfigure}
\begin{subfigure}[b]{0.49\textwidth}
\centering
\includegraphics[width=\textwidth]{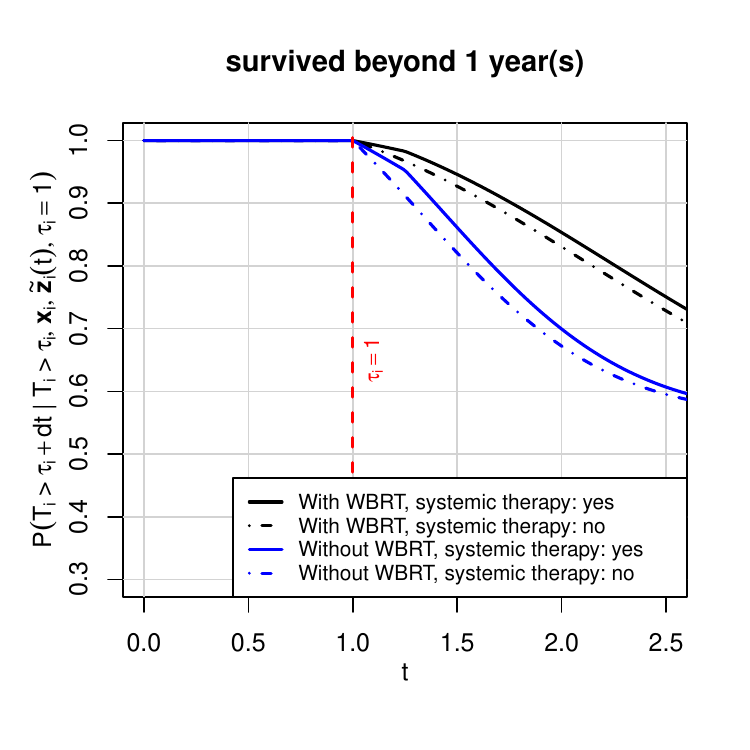}
\end{subfigure}
\hfill
\begin{subfigure}[b]{0.49\textwidth}
\centering
\includegraphics[width=\textwidth]{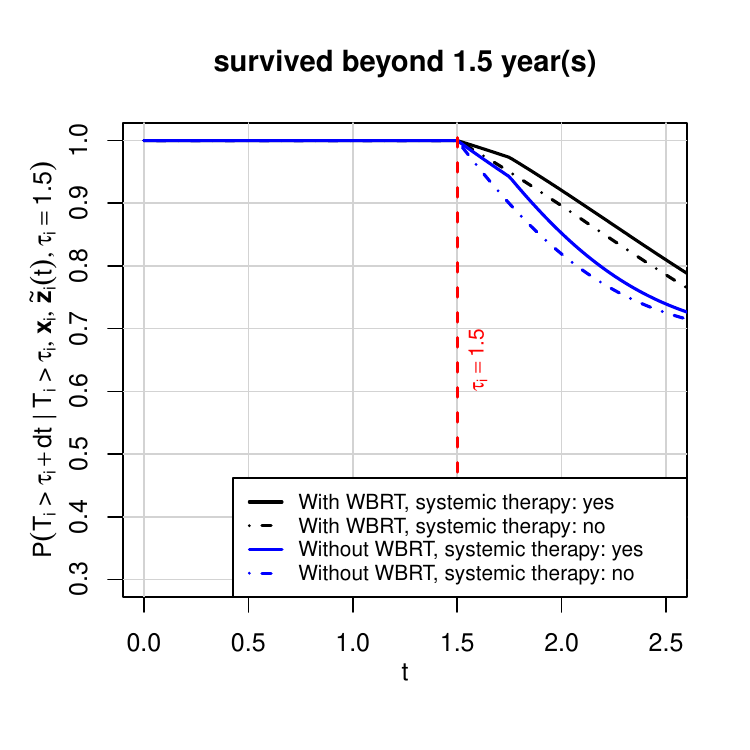}
\end{subfigure}
\caption{Predicted effect of the time-varying systemic therapy (either assumed to be applied such that $\boldsymbol{z}_i(\tau_i) = 1$, or not applied with $\boldsymbol{z}_i(\tau_i) = 0$ from $t = \tau_i$ for up to three months). See Equation \eqref{eq:dynamicPred} on computing brain metastasis-free survival for patients who have survived beyond $t = \tau_i$ with or without WBRT, for $\tau_i = \{0, 0.5, 1, 1.5\}$; adjusted for other time-fixed covariates: gender as female, number of MBM as $>1$, extracranial disease as present and age as its median value (63.3).
}
\label{fig:dynamicPrediction20241030_combine}
\end{figure}

Figure \ref{fig:dynamicPrediction20241030_combine} exhibits these predictive probabilities. Across all plots, patients who receive systemic therapy at $\tau_i$ demonstrate higher survival probabilities compared to those who do not. In addition, among patients who receive systemic therapy at $\tau_i$, those who have undergone WBRT show higher predictive survival probabilities than those who have not received WBRT. 

We may also evaluate when systemic therapy should be administered. Let us consider the predictive survival probabilities 6 months after systemic therapy has been first administered for varying values of $\tau_i$. From Figure \ref{fig:dynamicPrediction20241030_combine}, as the value of $\tau_i$ increases,  the predictive survival probabilities 6 months after the start of systemic therapy decrease. This indicates that systemic therapy has a diminished positive effect when initiated later in the study compared to earlier administration.

Thus, conditional probability estimates and predicted conditional survival plots, such as those shown in Figure \ref{fig:dynamicPrediction20241030_combine}, can provide valuable guidance to clinicians in determining whether and when to apply a sustained treatment for a patient.



\section{Discussion}\label{sec:discussion}
    Accelerated failure time models enable the regression analysis of survival data, where the failure times are either accelerated or decelerated by a multiplicative factor in the presence of several covariates. In this paper, we focused on semiparametric AFT models and provided a pathway to estimating such models with time-varying covariates under partly interval censoring, previously unavailable in the literature.
    
    In our approach, we selected Gaussian basis functions to approximate the unknown baseline hazard, which also help avoid numerical instabilities. A roughness penalty was introduced to construct a smooth approximation of the baseline hazard function. We then developed an algorithm that alternates between pseudo-Newton and MI steps to perform constrained optimisation. Combined with a marginal likelihood-based approach for automatic smoothing parameter selection, the algorithm successfully yielded the required maximum penalised likelihood estimates. The asymptotic results presented in Section \ref{sec:asymptotics} allow for statistical inference even in the presence of active constraints. In Section \ref{sec:simulation}, the simulation results  demonstrate that the proposed algorithm provides satisfactory performances. We also show that using a model that only accommodates right-censoring via midpoint imputation leads to larger biases and insufficient coverage probabilities. In Section \ref{sec:application}, we further illustrate the application of our approach using a real dataset on melanoma brain metastases. We also provide examples and plots of dynamic predictions that help assess the effects of the time-varying covariates considered in the model.

    In terms of computational times, each replication in the simulation study with a small sample size ($n = 100$) required 15 seconds on average to optimise. In contrast, the dataset from the WBRTMel trial ($n = 207$) required 3.5 minutes to optimise. For the simulation study with a large sample size ($n = 1000$), each replication took an average of 5 minutes to optimise. The \texttt{R} code used for the simulation studies and data analysis in this paper is publicly available at \url{https://github.com/aishwarya-b22/AFT_TVC_PIC}, where we provide a comprehensive vignette for fitting the model. In future work, we aim to reduce computational times using tools such as \texttt{Rcpp} to facilitate the implementation of accelerated failure time models.

    We also aim to develop robust model selection criteria to effectively compare and choose between AFT models and Cox proportional hazards models. Since both models are widely used in survival analysis but make different assumptions about the data, it is important to establish methods that guide practitioners in selecting the most appropriate model based on model fit and predictive performance. We plan to explore various approaches to identify model selection strategies that can be applied to a wide range of real-world datasets with diverse characteristics.

    The methodology proposed in the paper can also be readily extended to a joint AFT model that simultaneously captures both survival time and longitudinal predictors. The integration of longitudinal and survival data through joint models is particularly relevant in many observational medical studies and clinical trials, where repeated measurements of longitudinal biomarkers often show strong associations with time-to-event outcomes or overall survival. This approach specifically models the correlation between longitudinal data and time-to-event data, accounting for potential measurement error in the biomarkers.

\section*{Data Availability Statement}
    The WBRTMel trial dataset may be accessed upon reasonable request from the Melanoma Institute Australia.
    
\section*{Financial disclosure}

SNL is supported by Melanoma Institute Australia. 

\section*{Conflict of interest}

AB, DM, BL, SH and JM declare no potential conflict of interests. AMH has received an honorarium from Telix and fee for advisory board from Bayer. SNL has received fees for professional services from SkylineDx and a stipend for editorial services at the British Journal of Dermatology. 

\bibliographystyle{apalike}
\bibliography{ref_AFT}

\appendix
\input{supplementary}

\typeout{get arXiv to do 4 passes: Label(s) may have changed. Rerun}
	
\end{document}

%% file: supplementary.tex

\def\delE{\delta_i}
\def\delL{\delta^L_i}
\def\delR{\delta^R_i}
\def\delI{\delta^I_i}

\begin{center}
    Supplementary Materials for
\end{center}

\begin{center}
\Large{A maximum penalized likelihood approach for semiparametric AFT models with time-varying covariates and partly interval censoring}
\end{center}

\begin{center}
    by Aishwarya Bhaskaran, Ding Ma, Benoit Liquet, Angela Hong, Stephane Heritier,\newline Serigne N Lo and Jun Ma
\end{center}

\appendix 

\section{Vector and matrix representations for the gradients, Hessians and pseudo Hessians}

\subsection{Notation}
Let us define 
\begin{equation*}
		d_{\boldsymbol{\beta}}P(\boldsymbol{\beta}, \boldsymbol{\gamma}, \boldsymbol{\theta}), \quad D_{\boldsymbol{\beta}}P(\boldsymbol{\beta}, \boldsymbol{\gamma}, \boldsymbol{\theta}), \quad \nabla_{\boldsymbol{\beta}}P(\boldsymbol{\beta}, \boldsymbol{\gamma}, \boldsymbol{\theta})\quad\text{and}\quad H_{\boldsymbol{\beta}}P(\boldsymbol{\beta}, \boldsymbol{\gamma}, \boldsymbol{\theta}),
\end{equation*}
as the vector of differentials, derivative vector, gradient and Hessian of the penalized log-likelihood, $P(\boldsymbol{\beta}, \boldsymbol{\gamma}, \boldsymbol{\theta})$, with respect to $\boldsymbol{\beta}$. In addition, let
\begin{equation*}
	H^{-}_{\boldsymbol{\beta}}P(\boldsymbol{\beta}, \boldsymbol{\gamma}, \boldsymbol{\theta}),
\end{equation*}
denote the pseudo Hessian of the penalized log-likelihood with respect to $\boldsymbol{\beta}$. The pseudo Hessian is obtained from the original expression of the Hessian by only extracting the negative terms. 

The notation in this subsection applies to other model parameters as well.

\subsection{Expressions with respect to $\boldsymbol{\beta}$}
Firstly note that,
\begin{equation*}
	\begin{aligned}
		d_{\boldsymbol{\beta}}\kappa_i(y_i) 
		&= e^{-\boldsymbol{X}_i\boldsymbol{\beta}}\int_{0}^{y_i}e^{-z_i(s)\boldsymbol{\gamma}}ds\hspace{1mm}d(-\boldsymbol{X}_i\boldsymbol{\beta})\\
		&= -\kappa_i(y_i)\boldsymbol{X}_id\boldsymbol{\beta},
	\end{aligned}
\end{equation*}
and
\begin{equation*}
	\begin{aligned}
		d_{\boldsymbol{\beta}}\exp\{-\Lambda_0(\kappa_i(y_i))\}
		&= -\exp\{-\Lambda_0(\kappa_i(y_i))\}d\{\Lambda_0(\kappa_i(y_i))\}\\
		&= -\exp\{-\Lambda_0(\kappa_i(y_i))\}\lambda_0(\kappa_i(y_i))d\{\kappa_i(y_i)\}\\
		&=\exp\{-\Lambda_0(\kappa_i(y_i))\}\lambda_0(\kappa_i(y_i))\kappa_i(y_i)\boldsymbol{X}_id\boldsymbol{\beta}.
	\end{aligned}
\end{equation*}
\subsubsection{Gradient with respect to $\boldsymbol{\beta}$}
The gradient of $P(\boldsymbol{\beta}, \boldsymbol{\gamma}, \boldsymbol{\theta})$ with respect to $\boldsymbol{\beta}$ can be derived as,
\begin{equation*}
	\nabla_{\boldsymbol{\beta}}P(\boldsymbol{\beta}, \boldsymbol{\gamma}, \boldsymbol{\theta}) = D_{\boldsymbol{\beta}}P(\boldsymbol{\beta}, \boldsymbol{\gamma}, \boldsymbol{\theta})^T,
\end{equation*}
which can be shown to be,
\begin{normalsize}
	\begin{equation*}
		\begin{aligned}
			&D_{\boldsymbol{\beta}}P(\boldsymbol{\beta}, \boldsymbol{\gamma}, \boldsymbol{\theta})^T\\
			&= \sum_{i=1}^{n} \boldsymbol{X}_i^T \Bigg(\delE\left\{-\frac{\lambda_0^{'}(\kappa_i(y_i))\kappa_i(y_i)}{\lambda_0(\kappa_i(y_i))}- 1 + \lambda_0(\kappa_i(y_i))\kappa_i(y_i)\right\} + \delR\lambda_0(\kappa_i(y_i^L))\kappa_i(y_i^L)\Big)\\
			&\qquad -\delL\left[\frac{\exp\{-\Lambda_0(\kappa_i(y_i^R))\}\lambda_0(\kappa_i(y_i^R))\kappa_i(y_i^R)}{1 - \exp\{-\Lambda_0(\kappa_i(y_i^R))\}}\right]\\
			&\qquad +\delI\left[\frac{\exp\{-\Lambda_0(\kappa_i(y_i^L))\}\lambda_0(\kappa_i(y_i^L))\kappa_i(y_i^L) - \exp\{-\Lambda_0(\kappa_i(y_i^R))\}\lambda_0(\kappa_i(y_i^R))\kappa_i(y_i^R)}{\exp\{-\Lambda_0(\kappa_i(y_i^L))\} - \exp\{-\Lambda_0(\kappa_i(y_i^R))\}}\right]\Bigg).
		\end{aligned}
	\end{equation*}
\end{normalsize}
\normalsize

\subsubsection{Hessian with respect to $\boldsymbol{\beta}$}
Note that
\begin{equation*}
	H_{\boldsymbol{\beta}}P(\boldsymbol{\beta}, \boldsymbol{\gamma}, \boldsymbol{\theta}) = D_{\boldsymbol{\beta}}\left(D_{\boldsymbol{\beta}}P(\boldsymbol{\beta}, \boldsymbol{\gamma}, \boldsymbol{\theta})^T\right).
\end{equation*}
Then the Hessian of $P(\boldsymbol{\beta}, \boldsymbol{\gamma}, \boldsymbol{\theta})$ with respect to $\boldsymbol{\beta}$ can be shown to be,
\begin{scriptsize}
	\begin{equation*}
		\begin{aligned}
			&H_{\boldsymbol{\beta}}P(\boldsymbol{\beta}, \boldsymbol{\gamma}, \boldsymbol{\theta})\\
			&= \sum_{i=1}^{n}\boldsymbol{X}_i^T\Bigg[\delE\Bigg(\left[\frac{\lambda_0(\kappa_i(y_i))\lambda_0^{''}(\kappa_i(y_i))- \{\lambda_0^{'}(\kappa_i(y_i))\}^2}{\{\lambda_0(\kappa_i(y_i))\}^2} - \lambda_0^{'}(\kappa_i(y_i))\right]\{\kappa_i(y_i)\}^2+ \left\{\frac{\lambda_0^{'}(\kappa_i(y_i))}{\lambda_0(\kappa_i(y_i))} - \lambda_0(\kappa_i(y_i))\right\}\kappa_i(y_i)\Bigg)\\
			&\qquad - \delR\{\lambda_0^{'}(\kappa_i(y_i^L))\kappa_i(y_i^L) + \lambda_0(\kappa_i(y_i^L))\}\kappa_i(y_i^L) - \delL\Bigg\{\Bigg(\frac{\exp\{-\Lambda_0(\kappa_i(y_i^R))\}\left[\{\lambda_0(\kappa_i(y_i^R))\}^2 - \lambda_0^{'}(\kappa_i(y_i^R))\right]}{1-\exp\{-\Lambda_0(\kappa_i(y_i^R))\}}\\
			&\qquad \quad  + \left[\frac{\exp\{-\Lambda_0(\kappa_i(y_i^R))\}\lambda_0(\kappa_i(y_i^R))}{1-\exp\{-\Lambda_0(\kappa_i(y_i^R))\}}\right]^2\Bigg)\{\kappa_i(y_i^R)\}^2 - \left[\frac{\exp\{-\Lambda_0(\kappa_i(y_i^R))\}\lambda_0(\kappa_i(y_i^R))}{1-\exp\{-\Lambda_0(\kappa_i(y_i^R))\}}\right]\kappa_i(y_i^R)\Bigg\}\\
			&\qquad + \delI \Bigg\{\Bigg(\frac{\exp\{-\Lambda_0(\kappa_i(y_i^L))\}\left[\{\lambda_0(\kappa_i(y_i^L))\}^2 - \lambda_0^{'}(\kappa_i(y_i^L))\right]}{\exp\{-\Lambda_0(\kappa_i(y_i^L))\}-\exp\{-\Lambda_0(\kappa_i(y_i^R))\}} - \left[\frac{\exp\{-\Lambda_0(\kappa_i(y_i^L))\}\lambda_0(\kappa_i(y_i^L))}{\exp\{-\Lambda_0(\kappa_i(y_i^L))\}-\exp\{-\Lambda_0(\kappa_i(y_i^R))\}}\right]^2\Bigg)\{\kappa_i(y_i^L)\}^2\\
			&\qquad\quad  - \left[\frac{\exp\{-\Lambda_0(\kappa_i(y_i^L))\}\lambda_0(\kappa_i(y_i^L))}{\exp\{-\Lambda_0(\kappa_i(y_i^L))\}-\exp\{-\Lambda_0(\kappa_i(y_i^R))\}}\right]\kappa_i(y_i^L) -
			\Bigg(\frac{\exp\{-\Lambda_0(\kappa_i(y_i^R))\}\left[\{\lambda_0(\kappa_i(y_i^R))\}^2 - \lambda_0^{'}(\kappa_i(y_i^R))\right]}{\exp\{-\Lambda_0(\kappa_i(y_i^L))\}-\exp\{-\Lambda_0(\kappa_i(y_i^R))\}} \\
			&\qquad\quad + \left[\frac{\exp\{-\Lambda_0(\kappa_i(y_i^R))\}\lambda_0(\kappa_i(y_i^R))}{\exp\{-\Lambda_0(\kappa_i(y_i^L))\}-\exp\{-\Lambda_0(\kappa_i(y_i^R))\}}\right]^2\Bigg)\{\kappa_i(y_i^R)\}^2 + \left[\frac{\exp\{-\Lambda_0(\kappa_i(y_i^R))\}\lambda_0(\kappa_i(y_i^R))}{\exp\{-\Lambda_0(\kappa_i(y_i^L))\}-\exp\{-\Lambda_0(\kappa_i(y_i^R))\}}\right]\kappa_i(y_i^R)\\
			&\qquad\quad + 2\left(\frac{\exp\{-\Lambda_0(\kappa_i(y_i^L))\}\lambda_0(\kappa_i(y_i^L))\kappa_i(y_i^L)\exp\{-\Lambda_0(\kappa_i(y_i^R))\}\lambda_0(\kappa_i(y_i^R))\kappa_i(y_i^R)}{\left[\exp\{-\Lambda_0(\kappa_i(y_i^L))\}-\exp\{-\Lambda_0(\kappa_i(y_i^R))\}\right]^2}\right)\Bigg\}\Bigg]\boldsymbol{X}_i.
		\end{aligned}
	\end{equation*}
\end{scriptsize}
\normalsize

\subsubsection{Pseudo Hessian with respect to $\boldsymbol{\beta}$}
Subsequently, the pseudo Hessian of $P(\boldsymbol{\beta}, \boldsymbol{\gamma}, \boldsymbol{\theta})$ with respect to $\boldsymbol{\beta}$ is,
\begin{scriptsize}
	\begin{equation*}
		\begin{aligned}
			&H^{-}_{\boldsymbol{\beta}}P(\boldsymbol{\beta}, \boldsymbol{\gamma}, \boldsymbol{\theta})\\
			&= -\sum_{i=1}^{n}\boldsymbol{X}_i^T\Bigg[\delE\Bigg[\left\{\frac{ \lambda_0^{'}(\kappa_i(y_i))\kappa_i(y_i)}{\lambda_0(\kappa_i(y_i))} \right\}^2 + \lambda_0(\kappa_i(y_i))\kappa_i(y_i)\Bigg]  + \delR\lambda_0(\kappa_i(y_i^L))\kappa_i(y_i^L)\\
			&\qquad + \delL\Bigg\{\Bigg(\frac{\exp\{-\Lambda_0(\kappa_i(y_i^R))\}\{\lambda_0(\kappa_i(y_i^R))\}^2}{1-\exp\{-\Lambda_0(\kappa_i(y_i^R))\}} + \left[\frac{\exp\{-\Lambda_0(\kappa_i(y_i^R))\}\lambda_0(\kappa_i(y_i^R))}{1-\exp\{-\Lambda_0(\kappa_i(y_i^R))\}}\right]^2\Bigg)\{\kappa_i(y_i^R)\}^2\Bigg\}\\
			&\qquad + \delI \Bigg\{ \left[\frac{\exp\{-\Lambda_0(\kappa_i(y_i^L))\}\lambda_0(\kappa_i(y_i^L))}{\exp\{-\Lambda_0(\kappa_i(y_i^L))\}-\exp\{-\Lambda_0(\kappa_i(y_i^R))\}}\right]^2\{\kappa_i(y_i^L)\}^2 + \left[\frac{\exp\{-\Lambda_0(\kappa_i(y_i^L))\}\lambda_0(\kappa_i(y_i^L))}{\exp\{-\Lambda_0(\kappa_i(y_i^L))\}-\exp\{-\Lambda_0(\kappa_i(y_i^R))\}}\right]\kappa_i(y_i^L)\\
			&\qquad\quad  +
			\Bigg(\frac{\exp\{-\Lambda_0(\kappa_i(y_i^R))\}\{\lambda_0(\kappa_i(y_i^R))\}^2}{\exp\{-\Lambda_0(\kappa_i(y_i^L))\}-\exp\{-\Lambda_0(\kappa_i(y_i^R))\}} + \left[\frac{\exp\{-\Lambda_0(\kappa_i(y_i^R))\}\lambda_0(\kappa_i(y_i^R))}{\exp\{-\Lambda_0(\kappa_i(y_i^L))\}-\exp\{-\Lambda_0(\kappa_i(y_i^R))\}}\right]^2\Bigg)\{\kappa_i(y_i^R)\}^2\Bigg]\boldsymbol{X}_i.
		\end{aligned}
	\end{equation*}
\end{scriptsize}
\normalsize

\subsection{Expressions with respect to $\boldsymbol{\gamma}$}
Next note that,
\begin{equation*}
	\begin{aligned}
		d_{\boldsymbol{\gamma}}\kappa_i(y_i) 
		&= e^{-\boldsymbol{X}_i\boldsymbol{\beta}}\int_{0}^{y_i}e^{-z_i(s)\boldsymbol{\gamma}}d(-z_i(s)\boldsymbol{\gamma})ds\\
		&= -e^{-\boldsymbol{X}_i\boldsymbol{\beta}}\int_{0}^{y_i}e^{-z_i(s)\boldsymbol{\gamma}}z_i(s) ds d\boldsymbol{\gamma},\\
		D_{\boldsymbol{\gamma}}\kappa_i(y_i) 
		&= -e^{-\boldsymbol{X}_i\boldsymbol{\beta}}\int_{0}^{y_i}e^{-z_i(s)\boldsymbol{\gamma}}z_i(s)ds,
	\end{aligned}
\end{equation*}
and
\begin{equation*}
	\begin{aligned}
		D_{\boldsymbol{\gamma}}\{D_{\boldsymbol{\gamma}}\kappa_i(y_i)^T\}= e^{-\boldsymbol{X}_i\boldsymbol{\beta}}\int_{0}^{y_i}e^{-z_i(s)\boldsymbol{\gamma}}z_i(s)^Tz_i(s)ds.
	\end{aligned}
\end{equation*}
Also note that,
\begin{equation*}
	\begin{aligned}
		d\exp\{-\Lambda_0(\kappa_i(y_i))\}
		= -\exp\{-\Lambda_0(\kappa_i(y_i))\}\lambda_0(\kappa_i(y_i))d_{\boldsymbol{\gamma}}\kappa_i(y_i) .
	\end{aligned}
\end{equation*}

\subsubsection{Gradient with respect to $\boldsymbol{\gamma}$}
The gradient of $P(\boldsymbol{\beta}, \boldsymbol{\gamma}, \boldsymbol{\theta})$ with respect to $\boldsymbol{\gamma}$ can be derived as,
\begin{equation*}
	\nabla_{\boldsymbol{\gamma}}P(\boldsymbol{\beta}, \boldsymbol{\gamma}, \boldsymbol{\theta}) = D_{\boldsymbol{\gamma}}P(\boldsymbol{\beta}, \boldsymbol{\gamma}, \boldsymbol{\theta}) ^T\,
\end{equation*}
which can be shown to be,
\begin{small}
	\begin{equation*}
		\begin{aligned}
		& D_{\boldsymbol{\gamma}}P(\boldsymbol{\beta}, \boldsymbol{\gamma}, \boldsymbol{\theta}) ^T\\
			&= \sum_{i=1}^{n}\Bigg(-\delE z_i(y_i)^T + \delE \left\{\frac{\lambda_0^{'}(\kappa_i(y_i))}{\lambda_0(\kappa_i(y_i))} - \lambda_0(\kappa_i(y_i))\right\}D_{\boldsymbol{\gamma}}\kappa_i(y_i)^T - \delR\lambda_0(\kappa_i(y_i^L))D_{\boldsymbol{\gamma}}\kappa_i(y_i^L)^T\\
			&\qquad + \delL\left[\frac{\exp\{-\Lambda_0(\kappa_i(y_i^R))\}\lambda_0(\kappa_i(y_i^R))}{1 - \exp\{-\Lambda_0(\kappa_i(y_i^R))\}}\right]D_{\boldsymbol{\gamma}}\kappa_i(y_i^R)^T\\
			&\qquad + \delI\left[\frac{-\exp\{-\Lambda_0(\kappa_i(y_i^L))\}\lambda_0(\kappa_i(y_i^L))D_{\boldsymbol{\gamma}}\kappa_i(y_i^L)^T+\exp\{-\Lambda_0(\kappa_i(y_i^R))\}\lambda_0(\kappa_i(y_i^R))D_{\boldsymbol{\gamma}}\kappa_i(y_i^R)^T}{\exp\{-\Lambda_0(\kappa_i(y_i^L)) - \exp\{-\Lambda_0(\kappa_i(y_i^R))\}}\right]\Bigg).
		\end{aligned}
	\end{equation*}
\end{small}
\normalsize

\subsubsection{Hessian with respect to $\boldsymbol{\gamma}$}
Similar to the computation of $H_{\boldsymbol{\beta}}P(\boldsymbol{\beta}, \boldsymbol{\gamma}, \boldsymbol{\theta})$, note that
\begin{equation*}
	H_{\boldsymbol{\gamma}}P(\boldsymbol{\beta}, \boldsymbol{\gamma}, \boldsymbol{\theta}) = D_{\boldsymbol{\gamma}}\left(D_{\boldsymbol{\gamma}}P(\boldsymbol{\beta}, \boldsymbol{\gamma}, \boldsymbol{\theta})^T\right).
\end{equation*}
Then the Hessian of $P(\boldsymbol{\beta}, \boldsymbol{\gamma}, \boldsymbol{\theta})$ with respect to $\boldsymbol{\gamma}$ can be shown to be,
\begin{scriptsize}
	\begin{equation*}
		\begin{aligned}
			&H_{\boldsymbol{\gamma}}P(\boldsymbol{\beta}, \boldsymbol{\gamma}, \boldsymbol{\theta})\\
			&= \sum_{i=1}^{n}\Bigg[\delE\Bigg(D_{\boldsymbol{\gamma}}\kappa_i(y_i)^T\left[\frac{\lambda_0(\kappa_i(y_i))\lambda_0^{''}(\kappa_i(y_i))- \{\lambda_0^{'}(\kappa_i(y_i))\}^2}{\{\lambda_0(\kappa_i(y_i))\}^2} - \lambda_0^{'}(\kappa_i(y_i))\right]D_{\boldsymbol{\gamma}}\kappa_i(y_i)\\
			&\qquad\quad + \left\{\frac{\lambda_0^{'}(\kappa_i(y_i))}{\lambda_0(\kappa_i(y_i))} - \lambda_0(\kappa_i(y_i))\right\}D\{D_{\boldsymbol{\gamma}}\kappa_i(y_i)^T\}\Bigg) - \delR\left[\lambda_0(\kappa_i(y_i^L))D\{D_{\boldsymbol{\gamma}}\kappa_i(y_i^L)^T\} + D_{\boldsymbol{\gamma}}\kappa_i(y_i^L)^T\lambda_0^{'}(\kappa_i(y_i^L))D_{\boldsymbol{\gamma}}\kappa_i(y_i^L)\right]\\
			&\qquad + \delL \Bigg\{\left[\frac{\exp\{-\Lambda_0(\kappa_i(y_i^R))\}\lambda_0(\kappa_i(y_i^R))}{1-\exp\{-\Lambda_0(\kappa_i(y_i^R))\}}\right]D\{D_{\boldsymbol{\gamma}}\kappa_i(y_i^R)^T\}\\
			&\qquad \quad+ D_{\boldsymbol{\gamma}}\kappa_i(y_i^R)^T\Bigg(\frac{\exp\{-\Lambda_0(\kappa_i(y_i^R))\}\left[\lambda_0^{'}(\kappa_i(y_i^R)) - \{\lambda_0(\kappa_i(y_i^R))\}^2\right]}{1-\exp\{-\Lambda_0(\kappa_i(y_i^R))\}} - \left[\frac{\exp\{-\Lambda_0(\kappa_i(y_i^R))\}\lambda_0(\kappa_i(y_i^R))}{1-\exp\{-\Lambda_0(\kappa_i(y_i^R))\}}\right]^2\Bigg)D_{\boldsymbol{\gamma}}\kappa_i(y_i^R) \Bigg\}\\
			&\qquad + \delI
			\Bigg\{D_{\boldsymbol{\gamma}}\kappa_i(y_i^L)^T\Bigg(\frac{\exp\{-\Lambda_0(\kappa_i(y_i^L))\}\left[ \{\lambda_0(\kappa_i(y_i^L))\}^2 - \lambda_0^{'}(\kappa_i(y_i^L))\right]}{\exp\{-\Lambda_0(\kappa_i(y_i^L))\} - \exp\{-\Lambda_0(\kappa_i(y_i^R))\}} - \left[\frac{\exp\{-\Lambda_0(\kappa_i(y_i^L))\}\lambda_0(\kappa_i(y_i^L))}{\exp\{-\Lambda_0(\kappa_i(y_i^L))\} - \exp\{-\Lambda_0(\kappa_i(y_i^R))\}}\right]^2\Bigg)\\ &\qquad \qquad \times D_{\boldsymbol{\gamma}}\kappa_i(y_i^L)
			- \left[\frac{\exp\{-\Lambda_0(\kappa_i(y_i^L))\}\lambda_0(\kappa_i(y_i^L))}{\exp\{-\Lambda_0(\kappa_i(y_i^L))\} - \exp\{-\Lambda_0(\kappa_i(y_i^R))\}}\right]D\{D_{\boldsymbol{\gamma}}\kappa_i(y_i^L)^T\}\\
			&\qquad\quad - D_{\boldsymbol{\gamma}}\kappa_i(y_i^R)^T\Bigg(\frac{\exp\{-\Lambda_0(\kappa_i(y_i^R))\}\left[ \{\lambda_0(\kappa_i(y_i^R))\}^2 - \lambda_0^{'}(\kappa_i(y_i^R))\right]}{\exp\{-\Lambda_0(\kappa_i(y_i^L))\} - \exp\{-\Lambda_0(\kappa_i(y_i^R))\}} + \left[\frac{\exp\{-\Lambda_0(\kappa_i(y_i^R))\}\lambda_0(\kappa_i(y_i^R))}{\exp\{-\Lambda_0(\kappa_i(y_i^L))\} - \exp\{-\Lambda_0(\kappa_i(y_i^R))\}}\right]^2\Bigg)\\ 
			&\qquad \qquad \times D_{\boldsymbol{\gamma}}\kappa_i(y_i^R)
			+ \left[\frac{\exp\{-\Lambda_0(\kappa_i(y_i^R))\}\lambda_0(\kappa_i(y_i^R))}{\exp\{-\Lambda_0(\kappa_i(y_i^L))\}- \exp\{-\Lambda_0(\kappa_i(y_i^R))\}}\right]D\{D_{\boldsymbol{\gamma}}\kappa_i(y_i^R)^T\}\\
			&\qquad \quad + D_{\boldsymbol{\gamma}}\kappa_i(y_i^L)^T\Bigg(\frac{\exp\{-\Lambda_0(\kappa_i(y_i^L))\}\lambda_0(\kappa_i(y_i^L))\exp\{-\Lambda_0(\kappa_i(y_i^R))\}\lambda_0(\kappa_i(y_i^R))}{\left[\exp\{-\Lambda_0(\kappa_i(y_i^L))\} - \exp\{-\Lambda_0(\kappa_i(y_i^R))\}\right]^2}\Bigg)D_{\boldsymbol{\gamma}}\kappa_i(y_i^R)\\
			&\qquad \quad + D_{\boldsymbol{\gamma}}\kappa_i(y_i^R)^T\Bigg(\frac{\exp\{-\Lambda_0(\kappa_i(y_i^R))\}\lambda_0(\kappa_i(y_i^R))\exp\{-\Lambda_0(\kappa_i(y_i^L))\}\lambda_0(\kappa_i(y_i^L))}{\left[\exp\{-\Lambda_0(\kappa_i(y_i^L))\} - \exp\{-\Lambda_0(\kappa_i(y_i^R))\}\right]^2}\Bigg)D_{\boldsymbol{\gamma}}\kappa_i(y_i^L)\Bigg\}\Bigg].
		\end{aligned}
	\end{equation*}
\end{scriptsize}
\normalsize

\subsubsection{Pseudo Hessian with respect to $\boldsymbol{\gamma}$}
Hence, the pseudo Hessian of $P(\boldsymbol{\beta}, \boldsymbol{\gamma}, \boldsymbol{\theta})$ with respect to $\boldsymbol{\gamma}$ can be derived as,
\begin{scriptsize}
	\begin{equation*}
		\begin{aligned}
			&H_{\boldsymbol{\gamma}}^{-}P(\boldsymbol{\beta}, \boldsymbol{\gamma}, \boldsymbol{\theta})\\
			&= -\sum_{i=1}^{n}\Bigg[\delE\Bigg(D_{\boldsymbol{\gamma}}\kappa_i(y_i)^T\left\{\frac{ \lambda_0^{'}(\kappa_i(y_i))}{\lambda_0(\kappa_i(y_i))}\right\}^2 D_{\boldsymbol{\gamma}}\kappa_i(y_i) + \lambda_0(\kappa_i(y_i))D\{D_{\boldsymbol{\gamma}}\kappa_i(y_i)^T\}\Bigg) - \delR\lambda_0(\kappa_i(y_i^L))D\{D_{\boldsymbol{\gamma}}\kappa_i(y_i^L)^T\}\\
			&\qquad + \delL \Bigg\{ D_{\boldsymbol{\gamma}}\kappa_i(y_i^R)^T\Bigg(\frac{\exp\{-\Lambda_0(\kappa_i(y_i^R))\} \{\lambda_0(\kappa_i(y_i^R))\}^2}{1-\exp\{-\Lambda_0(\kappa_i(y_i^R))\}} + \left[\frac{\exp\{-\Lambda_0(\kappa_i(y_i^R))\}\lambda_0(\kappa_i(y_i^R))}{1-\exp\{-\Lambda_0(\kappa_i(y_i^R))\}}\right]^2\Bigg)D_{\boldsymbol{\gamma}}\kappa_i(y_i^R) \Bigg\}\\
			&\qquad + \delI
			\Bigg\{D_{\boldsymbol{\gamma}}\kappa_i(y_i^L)^T \left[\frac{\exp\{-\Lambda_0(\kappa_i(y_i^L))\}\lambda_0(\kappa_i(y_i^L))}{\exp\{-\Lambda_0(\kappa_i(y_i^L))\} - \exp\{-\Lambda_0(\kappa_i(y_i^R))\}}\right]^2 D_{\boldsymbol{\gamma}}\kappa_i(y_i^L) \\
			&\qquad \quad + \left[\frac{\exp\{-\Lambda_0(\kappa_i(y_i^L))\}\lambda_0(\kappa_i(y_i^L))}{\exp\{-\Lambda_0(\kappa_i(y_i^L))\} - \exp\{-\Lambda_0(\kappa_i(y_i^R))\}}\right]D\{D_{\boldsymbol{\gamma}}\kappa_i(y_i^L)^T\}\\
			&\qquad\quad + D_{\boldsymbol{\gamma}}\kappa_i(y_i^R)^T\Bigg(\frac{\exp\{-\Lambda_0(\kappa_i(y_i^R))\} \{\lambda_0(\kappa_i(y_i^R))\}^2}{\exp\{-\Lambda_0(\kappa_i(y_i^L))\} - \exp\{-\Lambda_0(\kappa_i(y_i^R))\}} + \left[\frac{\exp\{-\Lambda_0(\kappa_i(y_i^R))\}\lambda_0(\kappa_i(y_i^R))}{\exp\{-\Lambda_0(\kappa_i(y_i^L))\} - \exp\{-\Lambda_0(\kappa_i(y_i^R))\}}\right]^2\Bigg)D_{\boldsymbol{\gamma}}\kappa_i(y_i^R)\Bigg\}\Bigg].
		\end{aligned}
	\end{equation*}
\end{scriptsize}
\normalsize
\subsection{Expressions with respect to $\boldsymbol{\theta}$}
Let
\begin{equation*}
	\begin{aligned}
		\boldsymbol{\psi}(\kappa_i(y_i)) = [\psi_1(\kappa_i(y_i)), \dots, \psi_m(\kappa_i(y_i))]^T
		\quad\text{and}\quad
		\boldsymbol{\Psi}(\kappa_i(y_i)) = [\Psi_1(\kappa_i(y_i)), \dots, \Psi_m(\kappa_i(y_i))]^T.
	\end{aligned}
\end{equation*}
Also note that,
\begin{equation*}
	\begin{aligned}
		d\lambda_0(\kappa_i(y_i)) = \boldsymbol{\psi}(\kappa_i(y_i))d\boldsymbol{\theta}
		\quad\text{and}\quad
		d\Lambda_0(\kappa_i(y_i)) = \boldsymbol{\Psi}(\kappa_i(y_i))d\boldsymbol{\theta}.
	\end{aligned}
\end{equation*}

\subsubsection{Gradient with respect to $\boldsymbol{\theta}$}
The gradient of $P(\boldsymbol{\beta}, \boldsymbol{\gamma}, \boldsymbol{\theta})$ with respect to $\boldsymbol{\gamma}$ can be derived as,
\begin{equation*}
	\nabla_{\boldsymbol{\theta }}P(\boldsymbol{\beta}, \boldsymbol{\gamma}, \boldsymbol{\theta}) = D_{\boldsymbol{\theta }}P(\boldsymbol{\beta}, \boldsymbol{\gamma}, \boldsymbol{\theta})^T\,
\end{equation*}
which can be shown to be,
\begin{small}
	\begin{equation*}
		\begin{aligned}
			& D_{\boldsymbol{\theta }}P(\boldsymbol{\beta}, \boldsymbol{\gamma}, \boldsymbol{\theta})^T\\
			&= \sum_{i=1}^n \Bigg(\delE \left\{\frac{1}{\lambda_0(\kappa_i(y_i))}\boldsymbol{\psi}(\kappa_i(y_i)) - \boldsymbol{\Psi}(\kappa_i(y_i))\right\} - \delR\boldsymbol{\Psi}(\kappa_i(y_i^L))+ \delL\left[\frac{\exp\{-\Lambda_0(\kappa_i(y_i^R))\}}{1-\exp\{-\Lambda_0(\kappa_i(y_i^R))\}}\right]\boldsymbol{\Psi}(\kappa_i(y_i))\\
			&\qquad + \delI\left[\frac{1}{\exp\{-\Lambda_0(\kappa_i(y_i^L))\} - \exp\{-\Lambda_0(\kappa_i(y_i^R))\}}\right]\Big[-\boldsymbol{\Psi}(\kappa_i(y_i^L)) \exp\{-\Lambda_0(\kappa_i(y_i^L))\}\\
			&\qquad\quad  + \boldsymbol{\Psi}(\kappa_i(y_i^R))\exp\{-\Lambda_0(\kappa_i(y_i^R))\}\Big]\Bigg) - 2h\boldsymbol{\theta}^T\boldsymbol{R}.
		\end{aligned}
	\end{equation*}
\end{small}
\normalsize

\subsubsection{Hessian with respect to $\boldsymbol{\theta}$}
We have that,
\begin{equation*}
	H_{\boldsymbol{\theta}}P(\boldsymbol{\beta}, \boldsymbol{\gamma}, \boldsymbol{\theta}) = D_{\boldsymbol{\theta}}\left(D_{\boldsymbol{\theta}}P(\boldsymbol{\beta}, \boldsymbol{\gamma}, \boldsymbol{\theta})^T\right).
\end{equation*}
Then the Hessian of $P(\boldsymbol{\beta}, \boldsymbol{\gamma}, \boldsymbol{\theta})$ with respect to $\boldsymbol{\theta}$ can be derived as,
\begin{footnotesize}
	\begin{equation*}
		\begin{aligned}
			&H_{\boldsymbol{\theta}}P(\boldsymbol{\beta}, \boldsymbol{\gamma}, \boldsymbol{\theta})\\
			&= \sum_{i=1}^{n} \Bigg \{-\delE \left[\frac{1}{\{\lambda_0(\kappa_i(y_i))\}^2}\right]\boldsymbol{\psi}(\kappa_i(y_i))\boldsymbol{\psi}(\kappa_i(y_i))^T - \delL\boldsymbol{\Psi}(\kappa_i(y_i^R))\left(\frac{\exp\{-\Lambda_0(\kappa_i(y_i^R))\}}{[1-\exp\{-\Lambda_0(\kappa_i(y_i^R))\}]^2}\right)\boldsymbol{\Psi}(\kappa_i(y_i^R))^T\\
			&\qquad +\delI \Bigg(\left[\frac{1}{\exp\{-\Lambda_0(\kappa_i(y_i^L))\} - \exp\{-\Lambda_0(\kappa_i(y_i^R))\}}\right]\Big[\exp\{-\Lambda_0(\kappa_i(y_i^L))\}\boldsymbol{\Psi}(\kappa_i(y_i^L))\boldsymbol{\Psi}(\kappa_i(y_i^L))^T\\
			&\qquad \quad- \exp\{-\Lambda_0(\kappa_i(y_i^R))\}\boldsymbol{\Psi}(\kappa_i(y_i^R))\boldsymbol{\Psi}(\kappa_i(y_i^R))^T\Big] - \left(\frac{1}{\left[\exp\{-\Lambda_0(\kappa_i(y_i^L))\} - \exp\{-\Lambda_0(\kappa_i(y_i^R))\}\right]^2}\right)\\
			&\qquad \qquad \times \left[-\exp\{-\Lambda_0(\kappa_i(y_i^L))\}\boldsymbol{\Psi}(\kappa_i(y_i^L)) + \exp\{-\Lambda_0(\kappa_i(y_i^R))\}\boldsymbol{\Psi}(\kappa_i(y_i^R))\right]\\
			&\qquad \qquad \times
			\left[-\exp\{-\Lambda_0(\kappa_i(y_i^L))\}\boldsymbol{\Psi}(\kappa_i(y_i^L)) + \exp\{-\Lambda_0(\kappa_i(y_i^R))\}\boldsymbol{\Psi}(\kappa_i(y_i^R))\right]^T
			\Bigg)
			\Bigg\} - 2h\boldsymbol{R}.
		\end{aligned} 
	\end{equation*}
\end{footnotesize}
\normalsize

\subsubsection{Expression for $S_{\theta}$}
From $\nabla_{\boldsymbol{\theta }}P(\boldsymbol{\beta}, \boldsymbol{\gamma}, \boldsymbol{\theta})$, $S_{\boldsymbol{\theta}}$ can be calculated and shown to be a diagonal matrix with elements
\begin{equation*}
	\frac{\theta_u}{\sum_{i=1}^{n}\left[\delta_i\Psi_u(\kappa_i(y_i)) + \delta_i^R\Psi_u(\kappa_i(y_i^L)) + \delta_i^I\left\{\frac{\Psi_u(\kappa_i(y_i^L))\exp(-\Lambda_0(\kappa_i(y_i^L)))}{\exp(-\Lambda_0(\kappa_i(y_i^L))) - \exp(-\Lambda_0(\kappa_i(y_i^R)))}\right\} \right]+ 2h[\boldsymbol{R}_u\boldsymbol{\theta}]^{+}},
\end{equation*} 
where $\boldsymbol{R}_u$ denotes the $u$-th row of matrix $\boldsymbol{R}$ and $[\boldsymbol{\eta}]^+ = \max\{0, \boldsymbol{\eta}\}$.

\subsection{Second derivative with respect to $\boldsymbol{\beta}$ and $\boldsymbol{\gamma}$}
The second derivative of $P(\boldsymbol{\beta}, \boldsymbol{\gamma}, \boldsymbol{\theta})$ with respect to $\boldsymbol{\beta}$ and $\boldsymbol{\gamma}$ is,
\begin{equation*}
	D_{\boldsymbol{\gamma}}\left(D_{\boldsymbol{\beta}}P(\boldsymbol{\beta}, \boldsymbol{\gamma}, \boldsymbol{\theta})^T\right),
\end{equation*}
which can be expressed as,
\begin{scriptsize}
	\begin{equation*}
		\begin{aligned}
			& D_{\boldsymbol{\gamma}}\left(D_{\boldsymbol{\beta}}P(\boldsymbol{\beta}, \boldsymbol{\gamma}, \boldsymbol{\theta})^T\right)\\
			&= \sum_{i=1}^{n} \boldsymbol{X}_i^T\Bigg\{ \delE\Bigg[- \frac{\left\{\lambda_0(\kappa_i(y_i))\lambda_0^{'}(\kappa_i(y_i)) + \lambda_0(\kappa_i(y_i))\lambda_0^{''}(\kappa_i(y_i))\kappa_i(y_i) - (\lambda_0^{'}(\kappa_i(y_i)))^2\kappa_i(y_i)\right\}}{\{\lambda_0(\kappa_i(y_i))\}^2}\\ 
			&\qquad\quad + \{\lambda_0
			(\kappa_i(y_i)) + \lambda_0^{'}(\kappa_i(y_i))\kappa_i(y_i)\}\Bigg]D_{\boldsymbol{\gamma}}\kappa_i(y_i)\Bigg\} + \delR\left\{\lambda_0^{'}(\kappa_i(y_i^L))\kappa_i(y_i^L) + \lambda_0(\kappa_i(y_i^L))\right\}D_{\boldsymbol{\gamma}}\kappa_i(y_i^L)\\
			&\qquad - \delL\Bigg(\frac{\exp\{-\Lambda_0(\kappa_i(y_i^R))\}\left\{ \lambda_0(\kappa_i(y_i^R)) + \lambda_0^{'}(\kappa_i(y_i^R))\kappa_i(y_i^R)\right\}D_{\boldsymbol{\gamma}}\kappa_i(y_i^R)}{1 - \exp\{-\Lambda_0(\kappa_i(y_i^R))\}}\\
			&\qquad\quad - \frac{\{\lambda_0(\kappa_i(y_i^R))\}^2\kappa_i(y_i^R)\exp\{\Lambda_0(\kappa_i(y_i^R))\}D_{\boldsymbol{\gamma}}\kappa_i(y_i^R)}{[1-\exp\{-\Lambda_0(\kappa_i(y_i^R))\}]^2}\Bigg)\\
			&\qquad + \delI\Bigg[\Bigg\{\frac{\exp\{-\Lambda_0(\kappa_i(y_i^L))\}\left(\kappa_i(y_i^L)[\lambda_0^{'}(\kappa_i(y_i^L)) - \{\lambda_0(\kappa_i(y_i^L))\}^2] + \lambda_0(\kappa_i(y_i^L))\right)}{\exp\{-\Lambda_0(\kappa_i(y_i^L))\} - \exp\{-\Lambda_0(\kappa_i(y_i^R))\}}\Bigg\}D_{\boldsymbol{\gamma}}\kappa_i(y_i^L)\\
			&\qquad\quad - \Bigg\{\frac{\exp\{-\Lambda_0(\kappa_i(y_i^R))\}\left(\kappa_i(y_i^R)[\lambda_0^{'}(\kappa_i(y_i^R)) - \{\lambda_0(\kappa_i(y_i^R))\}^2] + \lambda_0(\kappa_i(y_i^R))\right)}{\exp\{-\Lambda_0(\kappa_i(y_i^L))\} - \exp\{-\Lambda_0(\kappa_i(y_i^R))\}}\Bigg\}D_{\boldsymbol{\gamma}}\kappa_i(y_i^R)\\
			&\qquad\quad + \frac{[\lambda_0(\kappa_i(y_i^R))\exp\{-\Lambda_0(\kappa_i(y_i^R))\}D_{\boldsymbol{\gamma}}\kappa_i(y_i^R) - \lambda_0(\kappa_i(y_i^L))\exp\{-\Lambda_0(\kappa_i(y_i^L))\}D_{\boldsymbol{\gamma}}\kappa_i(y_i^L)]}{[\exp\{-\Lambda_0(\kappa_i(y_i^L))\} - \exp\{-\Lambda_0(\kappa_i(y_i^R))\}]^2}\\
			&\qquad\quad \times [\lambda_0(\kappa_i(y_i^R))\exp\{-\Lambda_0(\kappa_i(y_i^R))\}\kappa_i(y_i^R) - \lambda_0(\kappa_i(y_i^L))\exp\{-\Lambda_0(\kappa_i(y_i^L))\}\kappa_i(y_i^L)]\Bigg]\Bigg\}.
		\end{aligned}
	\end{equation*}
\end{scriptsize}

\subsection{Second derivative with respect to $\boldsymbol{\beta}$ and $\boldsymbol{\theta}$}
Next, the second derivative of $P(\boldsymbol{\beta}, \boldsymbol{\gamma}, \boldsymbol{\theta})$ with respect to $\boldsymbol{\beta}$ and $\boldsymbol{\theta}$ is,
\begin{equation*}
	D_{\boldsymbol{\theta}}\left(D_{\boldsymbol{\beta}}P(\boldsymbol{\beta}, \boldsymbol{\gamma}, \boldsymbol{\theta})^T\right),
\end{equation*}
which can be derived as,
\begin{scriptsize}
	\begin{equation*}
		\begin{aligned}
			& D_{\boldsymbol{\theta}}\left(D_{\boldsymbol{\beta}}P(\boldsymbol{\beta}, \boldsymbol{\gamma}, \boldsymbol{\theta})^T\right)\\
			& =\sum^n_{i=1}\boldsymbol{X}_i\Bigg\{-\delE\left[\frac{\lambda_0(\kappa_i(y_i))\psi^{'}(\kappa_i(y_i)) - \lambda_0^{'}(\kappa_i(y_i))\psi(\kappa_i(y_i))}{\{\lambda_0(\kappa_i(y_i))\}^2} - \psi(\kappa_i(y_i))\right]\kappa_i(y_i) +\delR\kappa_i(y_i^L)\psi(\kappa_i(y_i^L))\\
			&\qquad - \delL\left(\frac{\exp\{-\Lambda_0(\kappa_i(y_i^R))\}\{\psi(\kappa_i(y_i^R)) - \Psi(\kappa_i(y_i^R))\lambda_0(\kappa_i(y_i^R))\}}{1 - \exp\{-\Lambda_0(\kappa_i(y_i^R))\}} - \frac{[\exp\{-\Lambda_0(\kappa_i(y_i^R))\}]^2\lambda_0(\kappa_i(y_i^R))\Psi(\kappa_i(y_i^R))}{[1-\exp\{-\Lambda_0(\kappa_i(y_i^R))\}]^2}\right)\kappa_i(y_i^R)\\
			&\qquad + \delI  \Bigg(\frac{\exp\{-\Lambda_0(\kappa_i(y_i^L))\}\{\psi(\kappa_i(y_i^L)) - \Psi(\kappa_i(y_i^L))\lambda_0(\kappa_i(y_i^L))\}\kappa_i(y_i^L)}{\exp\{-\Lambda_0(\kappa_i(y_i^L))\} - \exp\{-\Lambda_0(\kappa_i(y_i^R))\}}\\
			&\qquad\quad - \frac{\exp\{-\Lambda_0(\kappa_i(y_i^R))\}\{\psi(\kappa_i(y_i^R)) \Psi(\kappa_i(y_i^R))\lambda_0(\kappa_i(y_i^R))\}\kappa_i(y_i^R)}{\exp\{-\Lambda_0(\kappa_i(y_i^L))\} - \exp\{-\Lambda_0(\kappa_i(y_i^R))\}}\\
			&\qquad\quad + \frac{[\lambda_0(\kappa_i(y_i^R))\exp\{-\Lambda_0(\kappa_i(y_i^R))\}\kappa_i(y_i^R) - \lambda_0(\kappa_i(y_i^L))\exp\{-\Lambda_0(\kappa_i(y_i^L))\}\kappa_i(y_i^L)]}{[\exp\{-\Lambda_0(\kappa_i(y_i^L))\} - \exp\{-\Lambda_0(\kappa_i(y_i^R))\}]^2}\\
			&\qquad\qquad \times [\Psi(\kappa_i(y_i^R))\exp\{-\Lambda_0(\kappa_i(y_i^R))\} - \Psi(\kappa_i(y_i^L))\exp\{-\Lambda_0(\kappa_i(y_i^L))\}]\Bigg)\Bigg\}.
		\end{aligned}
	\end{equation*}
\end{scriptsize}
\normalsize

\subsection{Second derivative with respect to $\boldsymbol{\gamma}$ and $\boldsymbol{\theta}$}
Finally, the second derivative of $P(\boldsymbol{\beta}, \boldsymbol{\gamma}, \boldsymbol{\theta})$ with respect to $\boldsymbol{\gamma}$ and $\boldsymbol{\theta}$ is,
\begin{equation*}
	D_{\boldsymbol{\theta}}\left(D_{\boldsymbol{\gamma}}P(\boldsymbol{\beta}, \boldsymbol{\gamma}, \boldsymbol{\theta})^T\right),
\end{equation*}
which can be derived as,
\begin{scriptsize}
	\begin{equation*}
		\begin{aligned}
			&D_{\boldsymbol{\theta}}\left(D_{\boldsymbol{\gamma}}P(\boldsymbol{\beta}, \boldsymbol{\gamma}, \boldsymbol{\theta})^T\right)\\
			&= \sum^n_{i=1} \Bigg\{ \delE D_{\boldsymbol{\gamma}}\kappa_i(y_i)^T \left[\frac{\lambda_0(\kappa_i(y_i))\psi^{'}(\kappa_i(y_i)) - \lambda_0^{'}(\kappa_i(y_i))\psi(\kappa_i(y_i))}{\{\lambda_0(\kappa_i(y_i))\}^2} - \psi(\kappa_i(y_i))\right] - \delR D_{\boldsymbol{\gamma}}\kappa_i(y_i^L)^T\psi(\kappa_i(y_i^L))\\
			&\qquad + \delL D_{\boldsymbol{\gamma}}\kappa_i(y_i^R)^T\Big(\frac{\exp\{-\Lambda_0(\kappa_i(y_i^R))\}\{\psi(\kappa_i(y_i^R)) - \Psi(\kappa_i(y_i^R))\lambda_0(\kappa_i(y_i^R))\}}{1 - \exp\{-\Lambda_0(\kappa_i(y_i^R))\}}\\
			&\qquad\quad - \frac{[\exp\{-\Lambda_0(\kappa_i(y_i^R))\}]^2\lambda_0(\kappa_i(y_i^R))\Psi(\kappa_i(y_i^R))}{[1 - \exp\{-\Lambda_0(\kappa_i(y_i^R))\}]^2}\Big)\\
			&\qquad + \delI \Big(\frac{\exp\{\Lambda_0(\kappa_i(y_i^R))\}\{\psi(\kappa_i(y_i^R)) - \Psi(\kappa_i(y_i^R))\lambda_0(\kappa_i(y_i^R))\}D_{\boldsymbol{\gamma}}\kappa_i(y_i^R)^T}{\exp\{-\Lambda_0(\kappa_i(y_i^L))\} - \exp\{-\Lambda_0(\kappa_i(y_i^R))\}}\\
			&\qquad\quad - \frac{\exp\{\Lambda_0(\kappa_i(y_i^L))\}\{\psi(\kappa_i(y_i^L)) - \Psi(\kappa_i(y_i^L))\lambda_0(\kappa_i(y_i^L))\}D_{\boldsymbol{\gamma}}\kappa_i(y_i^L)^T}{\exp\{-\Lambda_0(\kappa_i(y_i^L))\} - \exp\{-\Lambda_0(\kappa_i(y_i^R))\}}\\
			&\qquad\quad + \frac{[\lambda_0(\kappa_i(y_i^L))\exp\{-\Lambda_0(\kappa_i(y_i^L))\}D_{\boldsymbol{\gamma}}\kappa_i(y_i^L)^T - \lambda_0(\kappa_i(y_i^R))\exp\{-\Lambda_0(\kappa_i(y_i^R))\}D_{\boldsymbol{\gamma}}\kappa_i(y_i^R)^T]}{[\exp\{-\Lambda_0(\kappa_i(y_i^L))\} - \exp\{-\Lambda_0(\kappa_i(y_i^R))\}]^2}\\
			&\qquad\qquad \times [\Psi(\kappa_i(y_i^R))\exp\{-\Lambda_0(\kappa_i(y_i^R))\} - \Psi(\kappa_i(y_i^L))\exp\{-\Lambda_0(\kappa_i(y_i^L))\}]\Big)\Bigg\}.
		\end{aligned}
	\end{equation*}
\end{scriptsize}